\documentclass[%
reprint,
superscriptaddress,
nofootinbib,
amsmath,amssymb,
aps,
]{revtex4-2}

\usepackage{graphicx}
\usepackage{dcolumn}
\usepackage{bm}

\usepackage{multirow}

\usepackage{amsmath,amssymb,amsfonts}
\usepackage{graphicx}
\usepackage{color}
\usepackage{appendix}
\usepackage{amsthm}
\usepackage{tikz}

\usepackage{mleftright}
\usepackage{url}
\usepackage[most]{tcolorbox}

\usepackage{algorithm}
\usepackage{algpseudocode}
\algnewcommand\algorithmicinput{\textbf{Input:}}
\algnewcommand\algorithmicoutput{\textbf{Output:}}
\algnewcommand\Input{\item[\algorithmicinput]}
\algnewcommand\Output{\item[\algorithmicoutput]}

\usepackage{tikz}
\usetikzlibrary{shapes,snakes,positioning,automata,arrows.meta,calc,decorations.markings,math,arrows.meta}
\tikzstyle{vertex}=[circle, draw, inner sep=0pt, minimum size=6pt]
\usepackage{xcolor}

\usepackage[export]{adjustbox}

\usepackage{geometry}
\newtheorem{definition}{Definition}

\newcommand{\nn}{\nonumber}

\newcommand{\paperpanel}[3]{%
  \begin{minipage}[b]{#1}%
    \centering
    \includegraphics[width=\linewidth]{#2}\\
    \normalfont(#3)
  \end{minipage}%
}

\def\a{\alpha}
\def\b{\beta}

\def\t{\tau}

\def\p{\psi}

\def\<{\langle}
\def\>{\rangle}

\def\ha{{\hat{a}}}

\def\t2{{\tilde{1}}}

\usepackage[normalem]{ulem}

\newcommand\encircle[1]{%
  \tikz[baseline=(X.base)] 
    \node (X) [draw, shape=circle, inner sep=0] {\strut #1};}

\begin{document}

	\title{Algorithmic Design of Heralded Linear Optical Circuits
    for Multipartite Entanglement}

\author{Jaehee Kim}
\affiliation{SKKU Advanced Institute of Nanotechnology (SAINT), Sungkyunkwan University, Suwon 16419, Korea}

\author{Hon Wai Lau}
\affiliation{Okinawa Institute of Science and Technology Graduate University, Okinawa 904-0495, Japan}

\author{William J. Munro}
\affiliation{Okinawa Institute of Science and Technology Graduate University, Okinawa 904-0495, Japan}

\author{Joonsuk Huh}
\email{joonsukhuh@yonsei.ac.kr}
\affiliation{Department of Chemistry, Yonsei University, Seoul 03722, Republic of Korea}
\affiliation{Department of Quantum Information, Yonsei University, Incheon 21983, Republic of Korea}
\affiliation{Department of Computational Science and Engineering, Yonsei University, Seoul 03722, Republic of Korea}

\author{Seungbeom Chin}
\email{sbthesy@gmail.com}
\affiliation{Okinawa Institute of Science and Technology Graduate University, Okinawa 904-0495, Japan}

	
	\begin{abstract}
Heralded multipartite entanglement is a key resource for various quantum information tasks. However, designing linear optical circuits that generate specific target states is generally challenging due to the complexity of the required optical structures. Here we formulate the design of heralded photonic circuits as an algorithmic graph-search problem. 
 Our framework enables the automated construction and optimization of heralded photonic circuits by  substantially reducing the search space using the linear quantum graph (LQG) picture. 
Our strategy reconstructs circuit structures as graphs in the picture and identifies suitable graphs automatically. 
As a result, we design efficient schemes for a broad range of useful multipartite resource states,
 including hypergraph magic states, quantum error correcting codes, general three-qubit states and length-1 caterpillar graph states. 
 Our work establishes an algorithmic framework for the systematic discovery of heralded resource states, 
 laying the foundation for the automated design of increasingly complex multipartite entangled resources.
	\end{abstract}
	
	
	\maketitle

\section{Introduction}

Multipartite entanglement is a central resource in quantum information science, enabling applications ranging from quantum communication and distributed quantum networks to measurement-based quantum computation and quantum error correction~\cite{horodecki2009quantum,pan2012multiphoton,walter2016multipartite}. Consequently, the generation of genuine multipartite entangled (GME) states has become a fundamental objective across a wide range of quantum platforms. 
In photonic quantum information processing, heralded generation occupies a special role of  producing usable quantum resources~\cite{kok2007linear, papp2009characterization,barz2010heralded,li2021heralded,chin2024shortcut,chin2024heralded} (see Ref.~\cite{forbes2025heralded} for a recent review). In contrast to postselected schemes~\cite{bouwmeester1999observation,pan2001experimental}, where successful events are identified after destructively measuring all output photons, heralded protocols use ancillary measurements to certify state preparation while leaving the generated state intact. The resulting entangled resources can be preserved for subsequent quantum information processing tasks, making heralded generation a key ingredient for scalable photonic architectures.

Despite their advantages, heralded photonic schemes are in many cases not straightforward to construct. Successful state generation requires a precise interplay between multi-photon interference, ancillary measurements, and post-selection conditions. All of them must be carefully coordinated to produce the desired output state. As the complexity of the target resource increases, the number of possible circuit configurations grows rapidly, making manual design based on trial and error increasingly challenging.
Recent studies have demonstrated that designing quantum optical experiments can be automated~\cite{krenn2021conceptual,cervera2022design,ruiz2023digital,hartnett2026automated}. However, these methods have so far seen only limited applications to the design of general heralded photonic circuits.

In this work, we overcome these limitations by proposing an algorithmic protocol for the systematic discovery of heralded photonic circuits that generate multipartite entangled states. Our protocol is based on the linear quantum graph (LQG) picture introduced in ~\cite{chin2021graph,chin2024shortcut}. By abstracting heralding detections with linear operations into boson subtraction operators and representing them as graph elements, the LQG picture reformulates the search for heralded schemes as a structured graph-search problem. Guided by the symmetries of LQG graphs, several heralded schemes have been designed to generate, e.g., GHZ, W, caterpillar graph, and Dicke states~\cite{chin2024heralded,chin2024exponentially,chin2026efficient,kang2026heralded}. However, such analytic constructions have remained state-specific and become increasingly difficult to generalize to more intricate entanglement structures. 
Our current approach combines graph-theoretic restrictions with the systematic numerical enumeration of  graph solutions, enabling the simultaneous discovery of various heralded photonic circuits.

We construct a vast graph repository of candidate circuits that satisfy the necessary condition for generating GME states, from which we can identify a variety of useful heralded schemes.
Our results include schemes for hypergraph magic states, quantum error-correcting codes (QECCs), general three-qubit states, and the $N=4$ cluster state. Some graph solutions in our results exhibit patterns that can be generalized to arbitrary system sizes. Based on these patterns, we also present general schemes for the $[[2k,1,2]]$ loss-tolerant QECC and arbitrary length-1 caterpillar graph states.

Our framework transforms heralded circuit design from a state-specific construction problem into a computational discovery paradigm. This is enabled by the convenient feature of LQG picture that imposes structural constraints on entanglement-generating circuits and thereby substantially reduces the search space.  Rather than producing a single solution for a particular target state, it systematically generates a large reusable repository of candidate graph structures that can be explored numerically. Beyond the schemes presented here, the vast majority of the graphs remain unexplored and    have the potential to yield other useful quantum resources.

Our work is organized as follows: Sec.~\ref{sec:algorithmic_strategy} explains our algorithmic strategy to construct graph repositories and design heralded GME-generating schemes from it. We also summarize our numerical data. Sec.~\ref{sec:results} presents representative efficient heralded schemes that are obtained from the graph repository. We show how the graph solutions are directly translated into linear optical circuits in the algorithm. Sec.~\ref{sec:discussions} discusses the capabilities and limitations of the current approach and suggests directions for future work.

\section{Algorithmic Strategy}\label{sec:algorithmic_strategy}

In the LQG picture, heralding detectors in linear optics, which can be described as photon-subtraction operators, are represented as a special subset of graphs called \emph{effective perfect matching (EPM) graphs}~\cite{chin2024shortcut}.
Each EPM graph determines the corresponding photon subtraction operations and hence the final state. 
Conversely, a heralded linear-optical circuit can be constructed directly from the graph using the translation rules introduced in Ref.~\cite{chin2024heralded}.
As a result, the circuit-design problem becomes the task of searching for suitable EPM graphs in the LQG picture. A brief introduction to the LQG picture and EPM graphs is provided in Appendix~\ref{appendix:LQG}.
In this section, we describe our algorithmic strategy from graph enumeration to the search for solutions corresponding to specific target states and the construction of heralded linear optical circuits. We also present our numerical statistics of the enumerated graphs.  

\subsection{Algorithm strategy}

\begin{figure*}
    \centering
    \includegraphics[width=0.9\textwidth]{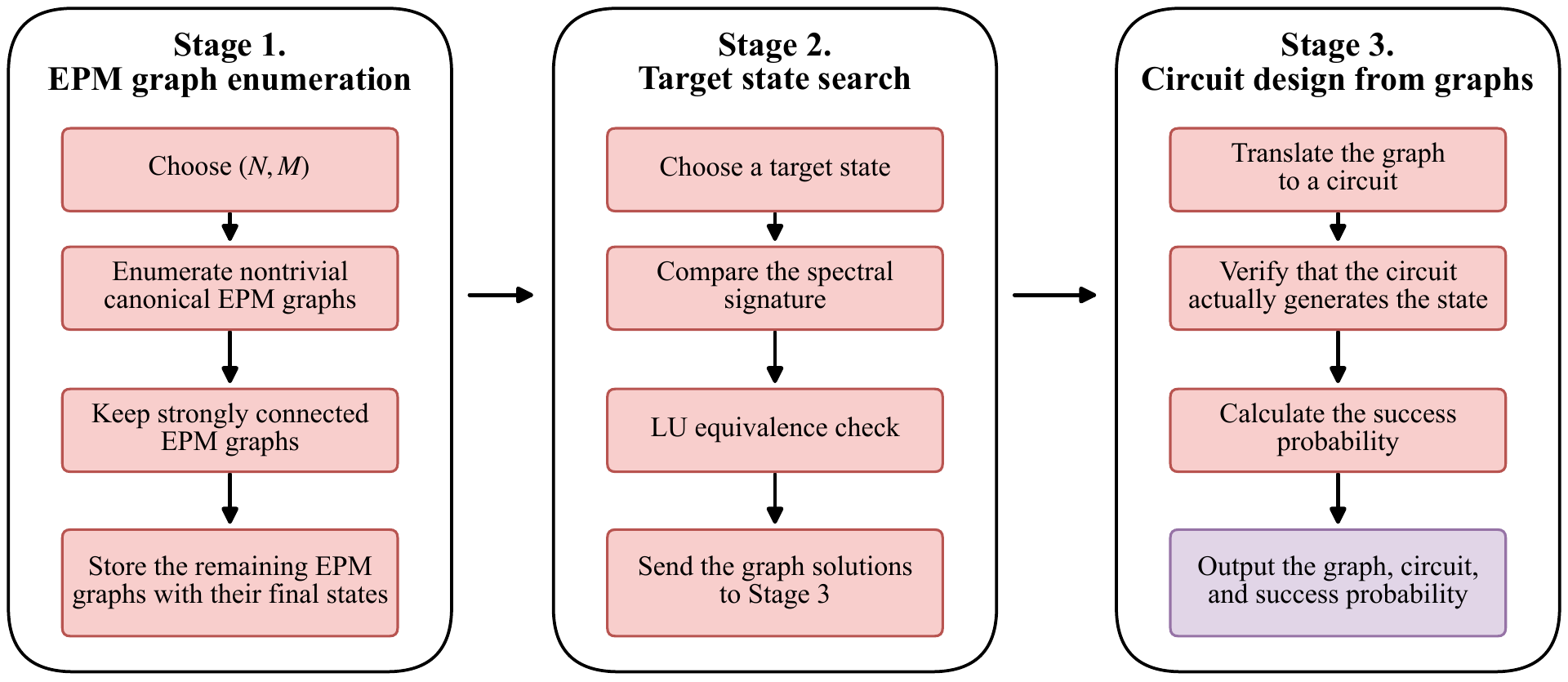}
    \caption{Summary of our algorithmic strategy. EPM graph repositories are constructed for different $(N,M)$ in Stage 1. Target states are selected and searched within the graph repositories in Stage 2.  The searched graphs are translated into heralded linear optical circuits using the LQG translation rules in Stage 3.}
    \label{fig:flowchart}
\end{figure*}

Our goal is to design heralded schemes for generating $N$-partite GME states with $M$ ancillary modes, which require $2N+M$ initial photons.
Our algorithmic approach for achieving the goal is summarized in Fig.~\ref{fig:flowchart} and consists of three main stages: EPM bigraph enumeration, target-state search, and translation from graph representations to linear optical circuits. 

$ $\\
\paragraph*{Stage 1. EPM graph enumeration.---} 
For fixed $N$ and $M$, the definition of EPM graphs and the necessary condition in Appendix~\ref{appendix:LQG} defines a finite number of EPM graphs that can be exhaustively enumerated numerically for fixed $N$ and $M$.

 For the algorithmic approach, we define the EPM graphs as follows: 

\begin{definition}[EPM Bipartite Graph]
\label{def:epm}
An EPM bipartite graph $G = (V, E, w)$ with node set $V = S \cup A \cup R$ consists of three disjoint node sets:
\begin{itemize}
  \item $S = \{S_0, S_1, \cdots, S_{N-1}\}$: System nodes; $\deg(S_i) = 2$, with one Red edge encoding $|0\rangle$ and one Blue edge encoding $|1\rangle$.
  \item $A = \{A_0, A_1, \cdots, A_{M-1}\}$: Ancilla nodes; $2\le \deg(A_j) \le N+M$, all edges Black encoding $|+\>$.
  \item $R = \{R_0, R_1, \ldots, R_{N+M-1}\}$: Photon-subtraction nodes; $|R| = |S| + |A|$.
\end{itemize}
All edges connect $S \cup A$ to $R$.
\end{definition}

Choosing a suitable number of ancilla nodes $M$ is crucial in enumerating EPM graphs. Larger $M$ increases the possibility for the graphs to generate useful state but also increases the computational cost exponentially by the inequality $2\le \deg(A_j) \le N+M$ $(0 \le j \le M-1)$ as we can  see in Tables~\ref{tab:enumeration} and \ref{tab:runtime_rss}.

For a fixed $N$ and $M$, we first enumerate all EPM graphs that satisfy the above definition and restrictions 1-3 in Appendix~\ref{appendix:stage1}. We use the graph canonization to store only non-isomorphic graphs  (Table~\ref{tab:enumeration}, non-trivial canonical graphs). Graph canonization identifies graphs that differ only by vertex relabeling, allowing redundant copies of the same graph structure to be removed.
We then eliminate graphs that are not strongly connected, since they do not satisfy the necessary condition to carry GME~(Appendix~\ref{appendix:stage1}, restriction 4). The strong-connectivity check reduces a substantial number of candidate graphs in our data (Table~\ref{tab:enumeration}, strongly connected graphs).
For each remaining graph, all perfect matchings (PMs) are enumerated and can be directly translated into the corresponding quantum state. We store the data as graphs with the generated final states in a repository.
Appendix~\ref{appendix:stage1} explains the technical details of the algorithm.

 At this stage, we treat the graphs as unweighted to reduce the runtime and memory. This implies that the corresponding amplitudes of those graphs are equally distributed among its connected dots, with no relative phase. The amplitude degree of freedom can be reintroduced in the next stage when searching for target states whose relative amplitudes are important.

\paragraph*{Stage 2. Target state search.---} In the second stage, we select an $N$-partite entangled target state and search the graph repository for EPM graphs that generate the state. 

Since we need to search for the graphs whose final states are local unitary (LU) equivalent to the target state, we first group the stored entries by the spectra of their reduced density matrices over all subsystems~\cite{nielsen2010quantum,kraus2010localprl,kraus2010localpra}. The spectra impose necessary conditions for two states to be LU equivalent (Definition~\ref{def:spectral_signature}). 
Then we select groups that have the same spectral signature as the target.

Next, we check whether the group actually contains the target. Since the identical spectral signature is not a sufficient condition for local LU equivalence~\cite{kraus2010localpra}, we search the group for states that are equivalent to the target up to permutations of spatial modes ($P$) and LU operations (Definition~\ref{def:pxlu}). We first check the relatively simple discrete operations $P$ and Pauli-$X$ operations $X$ ($P\times X$, Definition~\ref{def:px}). If we find graphs whose final states are transformed to the target by $P\times X$ operations, we send the graph data to Stage 3. Otherwise, we proceed to a more general $LU$ equivalence check numerically. If we find an optimized $U_0\otimes\cdots\otimes U_{N-1}$ operator that transforms the generated state to the target
we send the data to Stage 3. See Appendix~\ref{appendix:stage2} for more details. 

Additionally, when the target states have nontrivial relative amplitudes, a slightly different procedure is required to find graph solutions. We first select graphs that generate states that have the same superposed terms as the target and then check whether assigning edge weights can generate the actual target state. We can also search for graphs including redundant supports that cancel with each other using this procedure.

\paragraph*{Stage 3. Circuit design from graphs.---}
By identifying an EPM graph for a target state, we have now verified the photon subtraction operators that generate the state with fixed photon and mode numbers. We can directly build a heralded linear optical circuit using the graph as a blueprint for assembling optical elements following the LQG translation rules~(Ref.~\cite{chin2024heralded}, see Fig.~\ref{fig:translation_rules} of Appendix~\ref{appendix:LQG}). Each rule maps a local vertex structure to a linear optical component 
composed of multiport interferometers and particle-number-resolving detectors (PNRDs), which are connected according to the full graph topology. Our algorithm directly constructs the circuits, confirms they indeed generate the target states by operator evolutions, and also calculates the success probabilities. See Appendix~\ref{appendix:stage3} for more details.

\subsection{Statistics of Stage 1}

\begin{figure}[t]
    \centering
         \includegraphics[width=.5\textwidth]{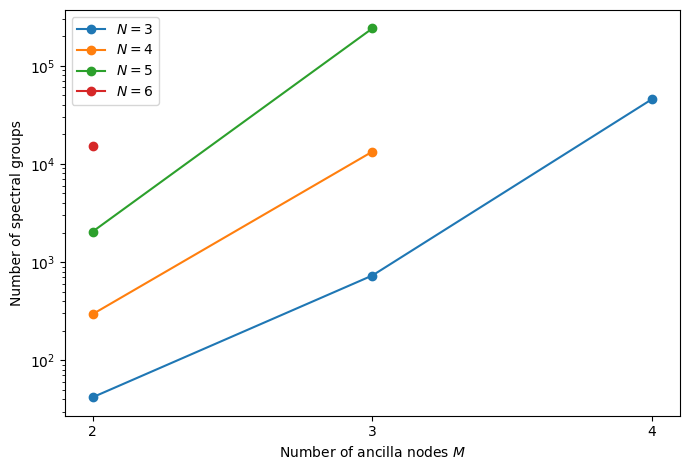}
        \caption{Number of distinct spectral-signature groups versus the number of ancilla nodes $M$ (from 2 to 4), for system sizes $N$ (from 3 to $6$) (see Table~\ref{tab:enumeration}, `spectra groups' column).}
        \label{fig:plot_groups}
\end{figure}

Our numerical statistics are summarized in Tables~\ref{tab:enumeration} and \ref{tab:runtime_rss}.
Table~\ref{tab:enumeration} reports the number of the enumerated graphs in Stage 1. The first column gives the number of enumerated  non-trivial canonical EPM graphs that satisfy Restrictions 1-3 in  Appendix~\ref{appendix:stage1}. The second column gives the number of strongly connected canonical EPM graphs.
We find that roughly one-third to nearly two-thirds of canonical graphs remain at this stage. Finally, the third column gives the number of different spectra groups in the resulting graph repository. Fig.~\ref{fig:plot_groups} plots the number of spectral groups for different $(N,M)$. The number of graphs increases more rapidly as $M$ increases than as  $N$ increases, as expected. 

Table~\ref{tab:runtime_rss} reports the corresponding benchmark
data for Stage 1, which consumes most of the computational resources: runtime and peak resident set size (RSS), which measures the physical memory used by the process.
The runtimes in Table~\ref{tab:runtime_rss} track the rapid growth of the search space: the raw candidate count of Eq.~\eqref{eq:general_ancilla_choice_count} grows from $\approx 4\times 10^{4}$ for $(3,2)$ to $\approx 2.5\times 10^{11}$ for $(5,3)$. Adding one ancilla node lengthens the enumeration time from $98~\mathrm{s}$ to about $2.5\times 10^{5}~\mathrm{s}$ for $N=3$ and from $363~\mathrm{s}$ to about $1.5\times 10^{6}~\mathrm{s}$ for $N=5$. Only one step further, the raw candidate count reaches $\approx 9\times 10^{12}$ for $(4,4)$ and $\approx 2\times 10^{15}$ for $(5,4)$, out of practical reach for the present implementation on a single desktop machine.


\begin{table}[t]
\caption{Numbers of enumerated EPM graphs. The last column counts the spectral-signature groups of the repository after the groups with the GHZ and W signatures are removed (Appendix~\ref{appendix:stage2}).}
\label{tab:enumeration}
\begin{ruledtabular}
\begin{tabular}{cccc}
$(N,M)$ & \begin{tabular}{@{}c@{}} non-trivial\\ canonical graphs \end{tabular} & \begin{tabular}{@{}c@{}} strongly connected\\graphs\end{tabular} & \begin{tabular}{@{}c@{}}spectra\\groups\end{tabular} 
\\
\hline
$(3,2)$ & 194 & 109 & $42$ \\
$(3,3)$ & $11{,}517$ & $6{,}644$ & $730$  \\
$(3,4)$ & $1{,}510{,}456$ & $948{,}274$ & $45{,}659$  \\
$(4,2)$ & $1{,}568$ & $693$ & $295$  \\
$(4,3)$ & $196{,}209$ & $94{,}361$ & $13{,}323$ \\
$(5,2)$ & $12{,}609$ & $4{,}647$ & $2{,}043$ \\
$(5,3)$ & $3{,}420{,}511$ & $1{,}391{,}872$ & $239{,}643$  \\
$(6,2)$ & $106{,}370$ & $32{,}493$ & $15{,}073$  \\
\end{tabular}
\end{ruledtabular}
\end{table}

\begin{table}[t]
\caption{Runtime and RSS for the enumeration data. All runs were performed on a single desktop machine (Apple M4 Pro, 12
cores, 64\,GB RAM) using Python~3.12 with NumPy~2.5, SciPy~1.18,
SymPy~1.14, QuTiP~5.3, and python-igraph~1.0. Runtime covers the graph enumeration; peak memory also covers the subsequent construction of the repository and its spectral-signature groups (see Appendix~\ref{appendix:methods}).}
\label{tab:runtime_rss}
\begin{ruledtabular}
\begin{tabular}{ccc}
$(N,M)$ & runtime (s) & peak RSS (GiB) \\
\hline
$(3,2)$ & $0.4$ & $0.024$ \\
$(3,3)$ & $98$ & $0.180$ \\
$(3,4)$ & $252{,}969$ & $24.122$ \\
$(4,2)$ & $7$ & $0.130$ \\
$(4,3)$ & $9{,}812$ & $2.601$ \\
$(5,2)$ & $363$ & $0.173$ \\
$(5,3)$ & $1{,}480{,}055$ &  $23.658$ \\
$(6,2)$ & $23{,}332$ & $0.980$ \\
\end{tabular}
\end{ruledtabular}
\end{table}


\section{Results}\label{sec:results}

Our graph repository contains a broad range of heralded schemes. In this section, we present the most useful schemes identified in our search. We focus on schemes that, to our knowledge, have not been previously reported or improve upon known schemes in terms of the required photon number and success probability. Our results include schemes for hypergraph magic states, quantum error-correcting codes, general three-qubit states, and the $N=4$ cluster state. Some graph structures in our search exhibit patterns that can be generalized to arbitrary system sizes. Based on these patterns, we present general schemes for $[[2k,1,2]]$ error correcting codes and arbitrary length-1 caterpillar graph states.

It is worth noting that graphs in $(N,M)$ repository require $2N +M$ photons to generate the expected target state. We present circuits of representative states from our graph solutions here. All the success probabilities below
assume feedforward single-qubit phase corrections (Appendix~\ref{appendix:stage3}), and all solutions achieve perfect fidelity. A complete set of circuit structures and their success probabilities are provided in Appendix~\ref{appendix:circuits}.

\subsection{Hypergraph magic states}

\begin{figure*}[t]
    \centering
        \paperpanel{0.315654\textwidth}{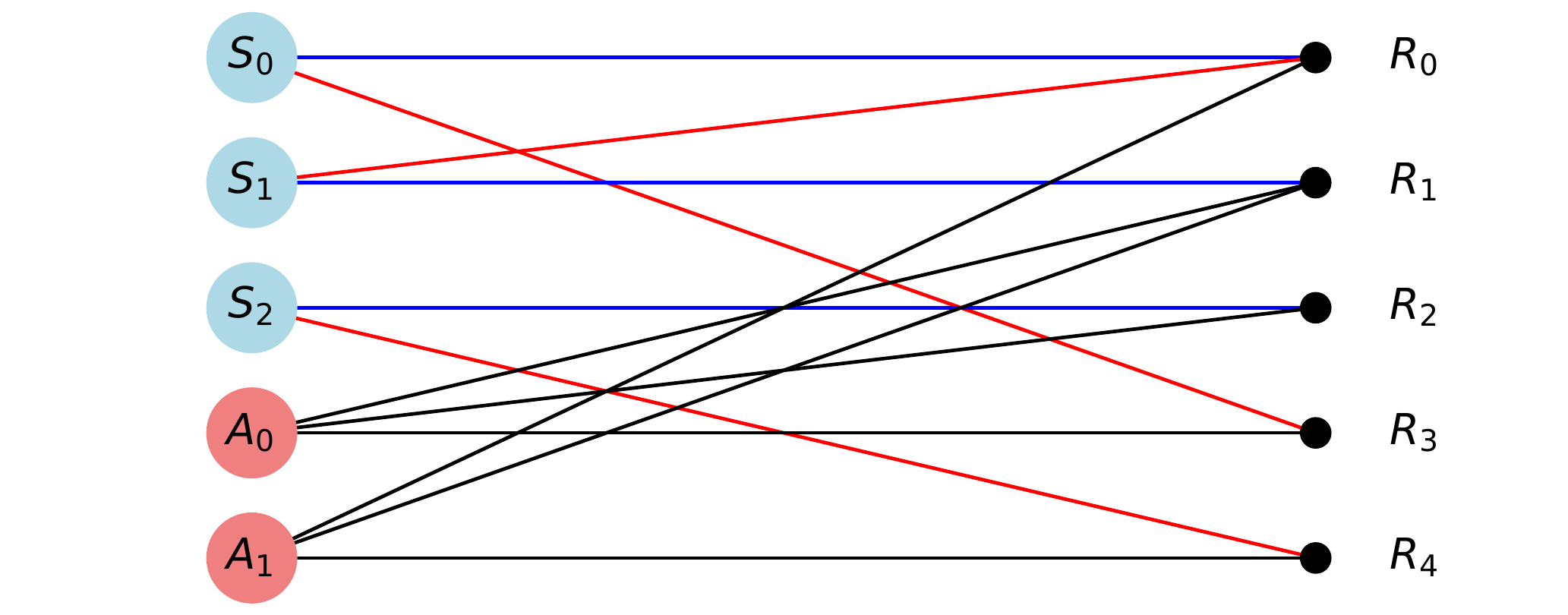}{a}\hfill
        \paperpanel{0.315654\textwidth}{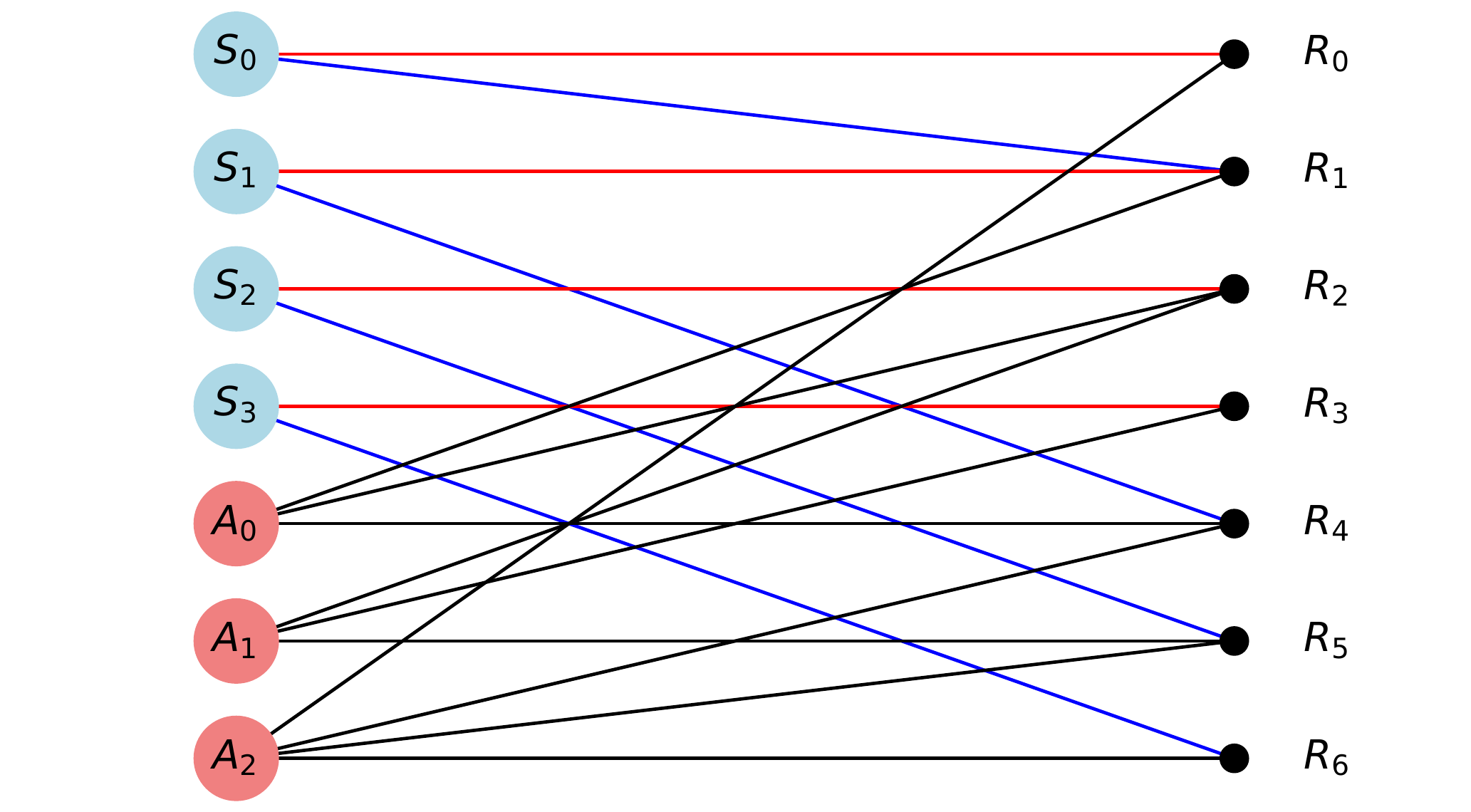}{b}\hfill
        \paperpanel{0.315654\textwidth}{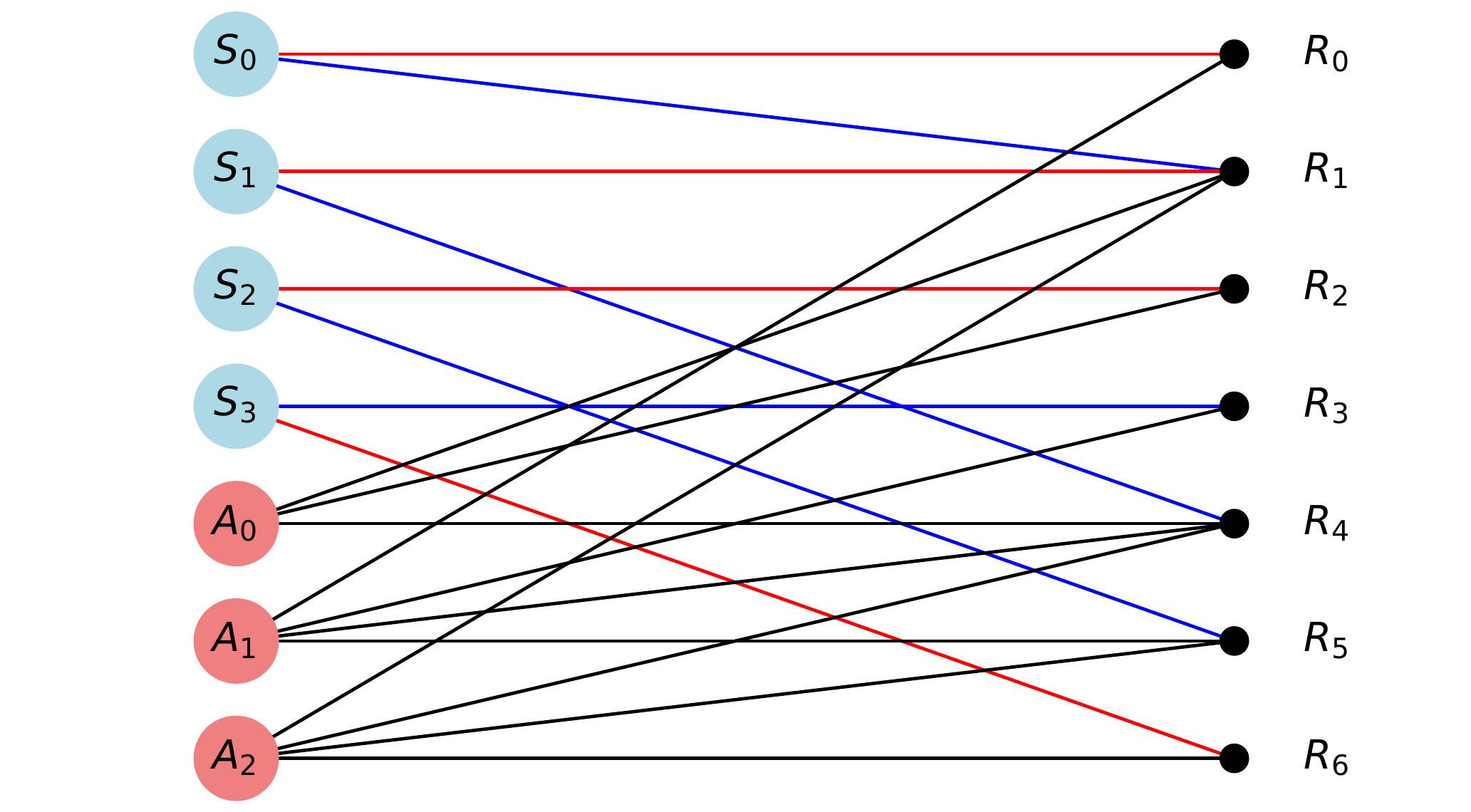}{c}
        \caption{EPM graph solutions for hypergraph magic states: (a) $H_0CCZ_{012}|+++\>$ for $N=3$, using 8 photons; and two solutions, (b) and (c), for $H_0CCCZ_{0123}|++++\>$ for $N=4$, each using 11 photons. 
        }
        \label{fig:magic_N}
\end{figure*} 

\begin{figure}[t]
    \centering
         \includegraphics[width=.483\textwidth]{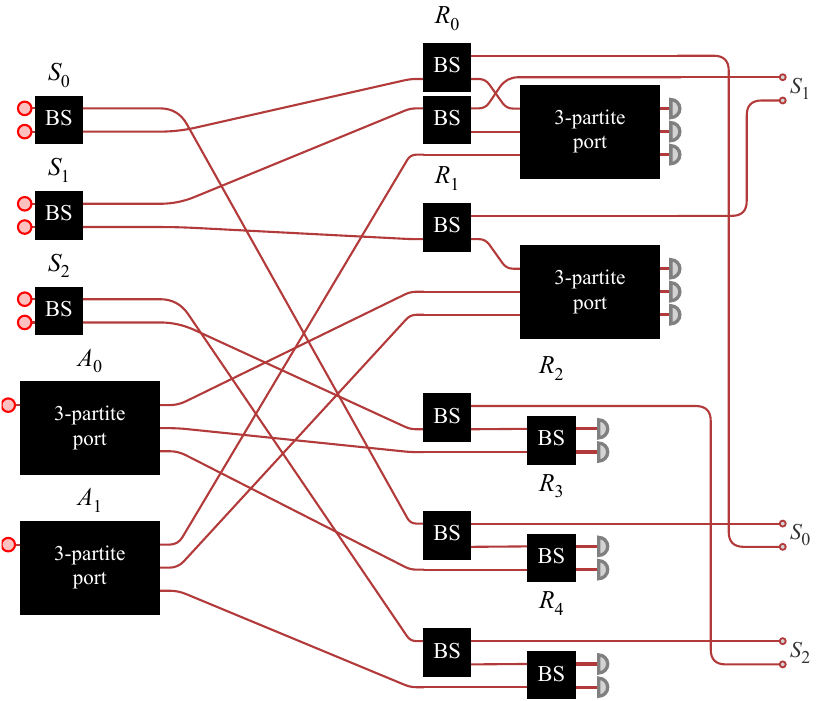}
        \caption{Linear optical circuit for generating $N=3$ hypergraph magic state, which is directly constructed through the translation rules from graphs to circuits.}
        \label{fig:magic3_circuit}
\end{figure} 

It is well-known that arbitrary stabilizer states can be generated by fusion gates~\cite{browne2005resource,varnava2006loss} up to local Clifford operations. In contrast, universal quantum computation requires magic resources, which lie outside the stabilizer formalism and therefore cannot be generated using fusion gates alone~\cite{bravyi2005universal,veitch2014resource}. Our graph solutions include particularly useful magic states, i.e., hypergraph magic states~\cite{rossi2013quantum,huang2024demonstration,poderini2026quantum} for $N=3$ and 4. 

\paragraph*{$N=3$ hypergraph magic state.---} The hypergraph state is  generated by applying non-Clifford $CCZ$ gates to $|+++\>$,
\begin{align}
 \text{CCZ}|+++\> \equiv |\text{CCZ}\>.    
\end{align}
While previously known schemes generate the state with the heralded CCZ gate in linear optics~\cite{ralph2007efficient,uskov2009maximal}, we can find an alternative heralded scheme in our $(3,2)$ repository that directly generates the target state with a higher success probability. 

The EPM graph solution is given  as Fig.~\ref{fig:magic_N} (a). 
The final state generated by this circuit is actually
\begin{align}\label{eq:n3_magic}
 H_0|CCZ\> = \frac{1}{2}(|000\> + |001\> + |010\> + |111\>). 
\end{align}
In Stage 3, the graph is translated into the linear optical circuit in Fig.~\ref{fig:magic3_circuit}.
In the dual-rail encoding, the Hadamard gate is realized by a local 50:50 beam splitter (BS), hence we can easily transform the state into $|CCZ\>$ deterministically.

Our scheme requires 8 photons and has a success probability of $P_{\text{succ}} = 1/128 \approx 7.8\times 10^{-3}$.  The previous best scheme requires a three-photon initial state and a heralded Toffoli gate with three ancillary photons, corresponding to 6 photons in total, but has a lower success probability of $P_{\text{succ}} \approx 3.4\times 10^{-3}$~\cite{uskov2009maximal}. Thus, our scheme improves the success probability at the cost of two additional photons.

\paragraph*{$N=4$ hypergraph magic states.---} The CCZ gate can be generalized to $C^{(N-1)}Z$ gates for arbitrary $N$-partite systems. Our $(4,3)$ repository includes two schemes for generating 
\begin{align}
  CCCZ_{0123}|++++\>\equiv |CCCZ\>,
\end{align} which are presented in Fig.~\ref{fig:magic_N}(b) and (c).
More exactly, the state generated by both schemes is
\begin{align}
&H_0CCCZ_{0123}|++++\> \nn \\
&= \frac{1}{2\sqrt{2}}(|0000\> +  |0100\> + |0010\> + |0001\> \nn \\
&~~~~~~~~~~~+|0110\> + |0011\> + |0101\>  + |1111\>).
\end{align}
This state directly enables the implementation of the four-qubit CCCZ gate in measurement-based quantum computation (MBQC).

The circuits corresponding to the graphs are presented in Appendix~\ref{appendix:hypergraph_4}. The success probabilities of the graphs in Fig.~\ref{fig:magic_N}(b) and (c) are approximately $3.51 \times 10^{-4}$ and $2.03\times 10^{-5}$, respectively. To our best knowledge, no heralded scheme for $|CCCZ\>$ has been reported previously.


\subsection{Quantum error correcting codes}

\begin{figure*}[t]
    \centering
    \paperpanel{0.255\textwidth}{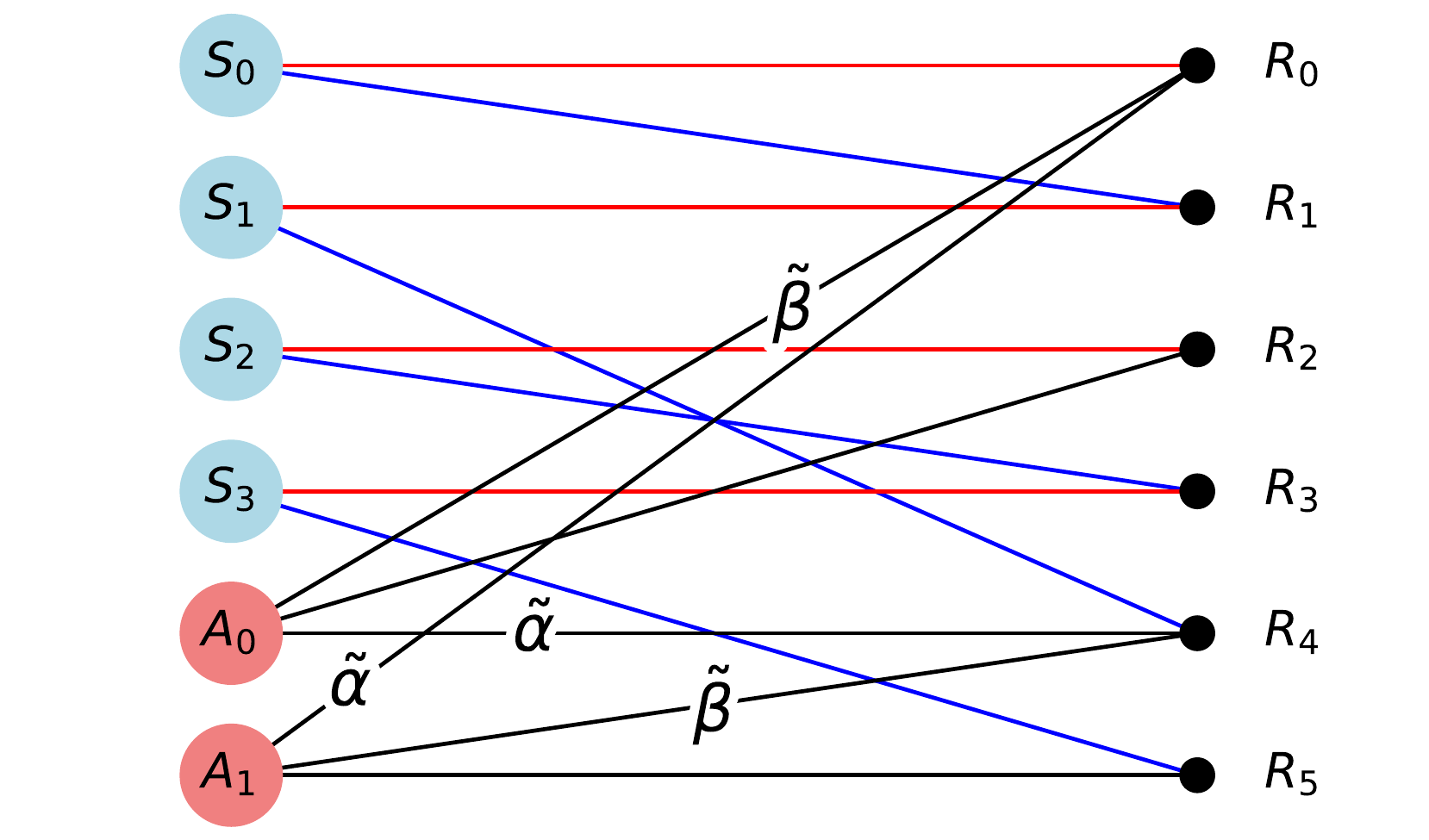}{a}\hfill
    \paperpanel{0.255\textwidth}{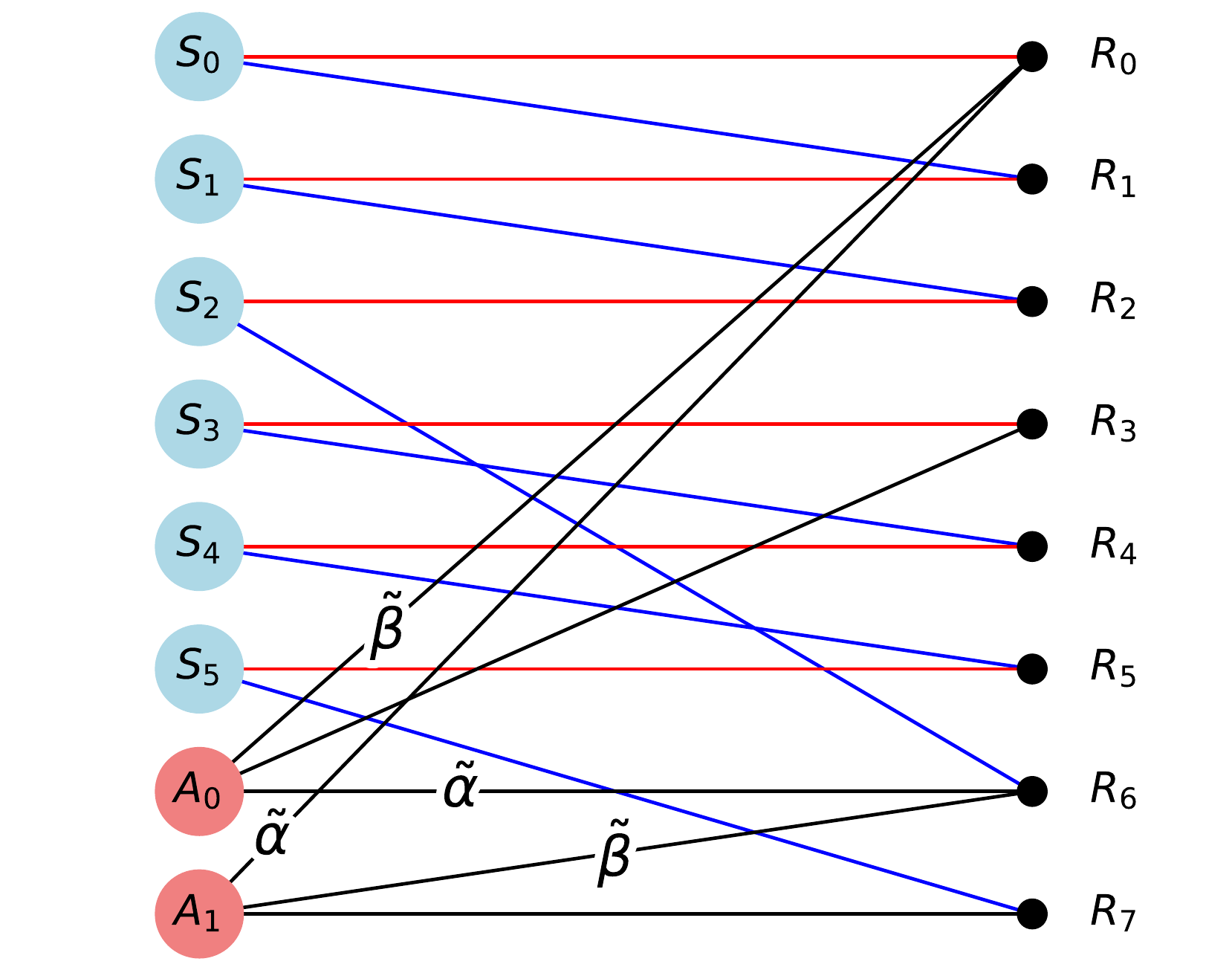}{b}\hfill
    \paperpanel{0.178\textwidth}{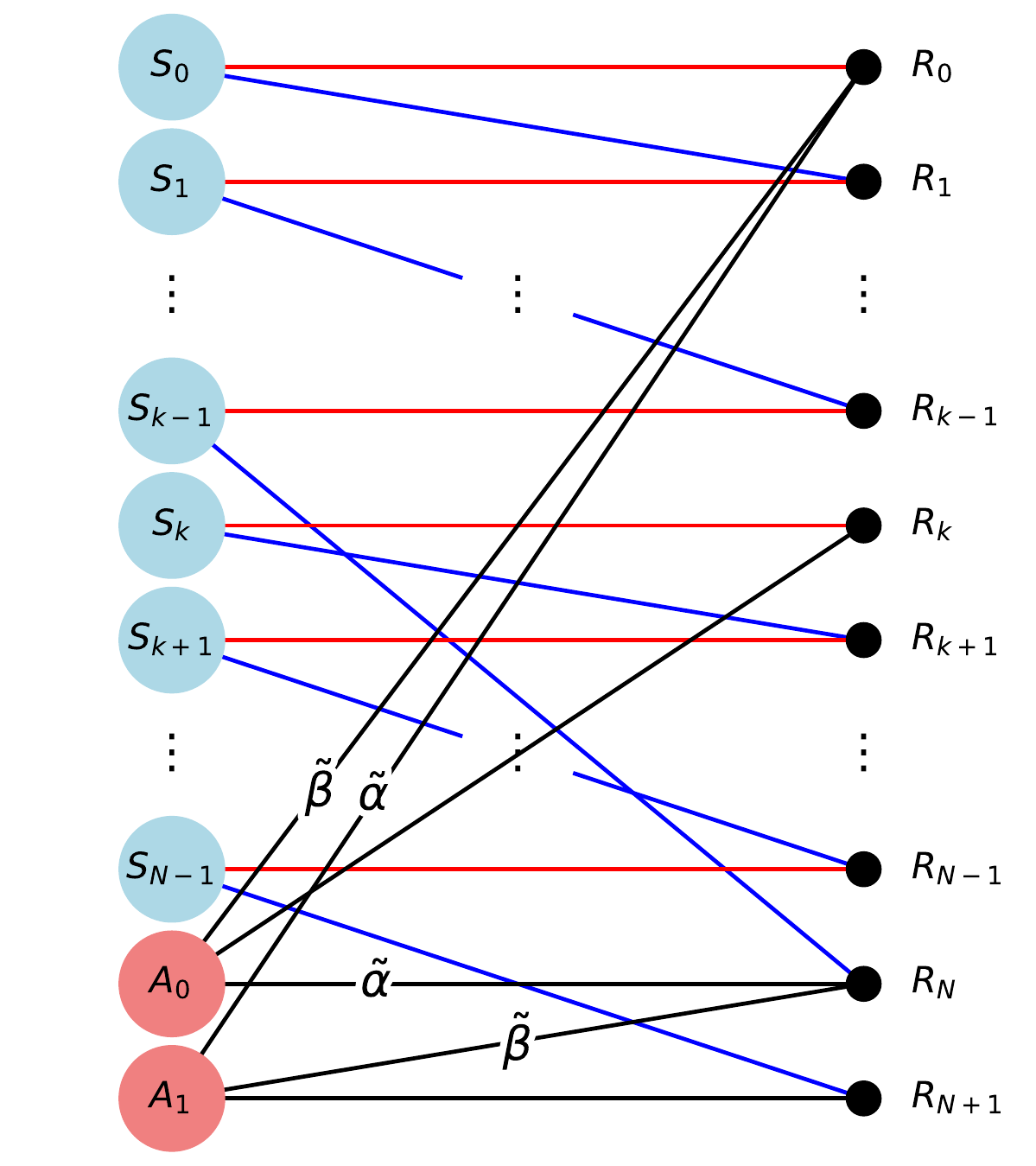}{c}\hfill
    \paperpanel{0.265\textwidth}{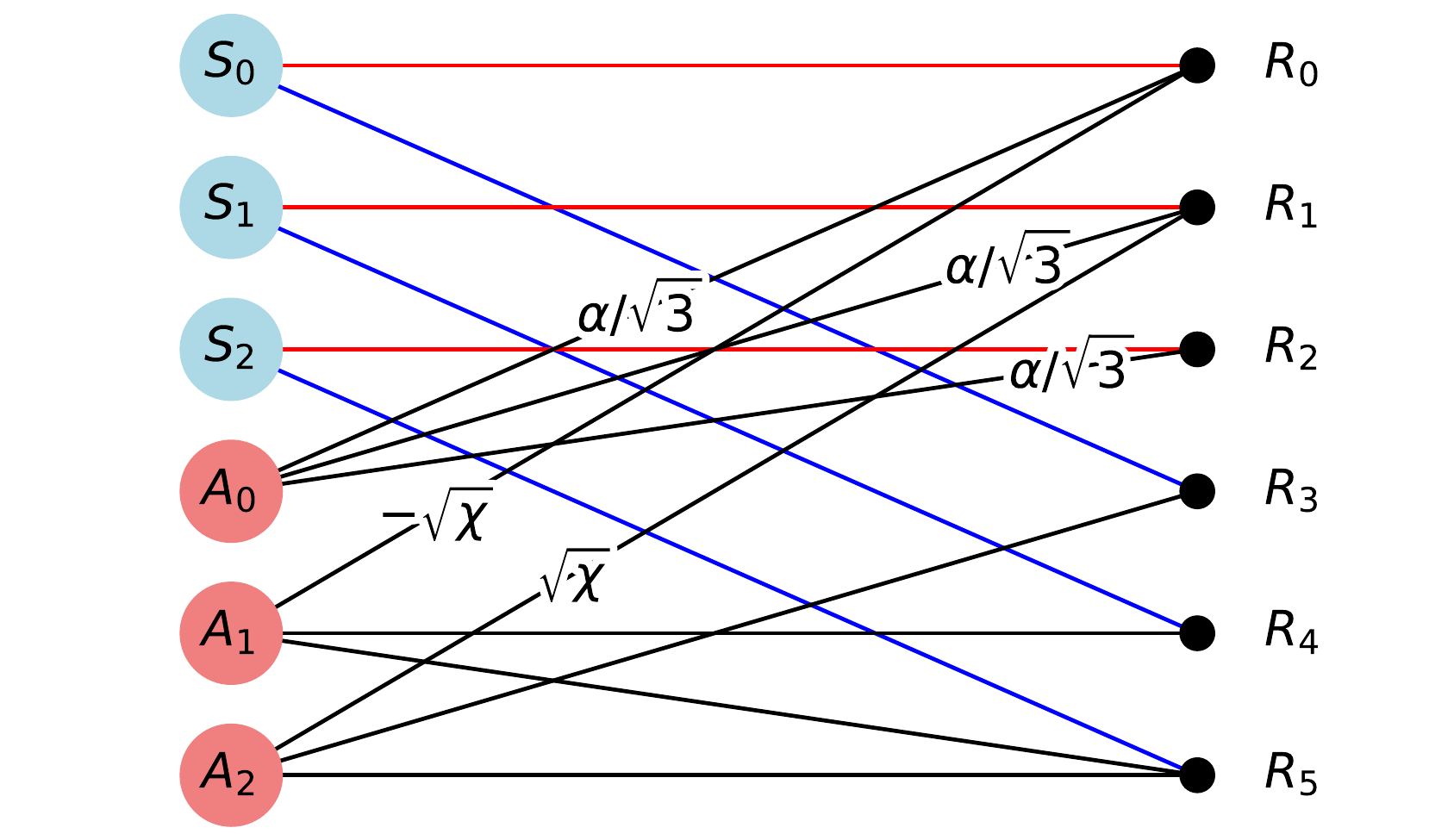}{d}
        \caption{(a), (b) Graph solutions for generating the $[[2k,1,2]]$ loss-tolerant error correcting codes for $k=2,3$, (c) an arbitrary $[[2k,1,2]]$ loss-tolerant error correcting code with $N=2k$, inferred from the patterns of the numerically generated graphs in (a) and (b). Only the free-parameter edges are labeled, with $\tilde{\alpha} \equiv \a/\sqrt{2}$ and $\tilde{\beta} \equiv \b/\sqrt{2}$. (d) The $N=3$ amplitude-damping code that has the highest success probability among 11 solutions in $(3,3)$ repository. The amplitude conventions of (d) are given in Appendix~\ref{appendix:N3_ADC}.}
        \label{fig:qec}
\end{figure*}

\begin{figure}[t]
    \centering
         \includegraphics[width=0.49\textwidth]{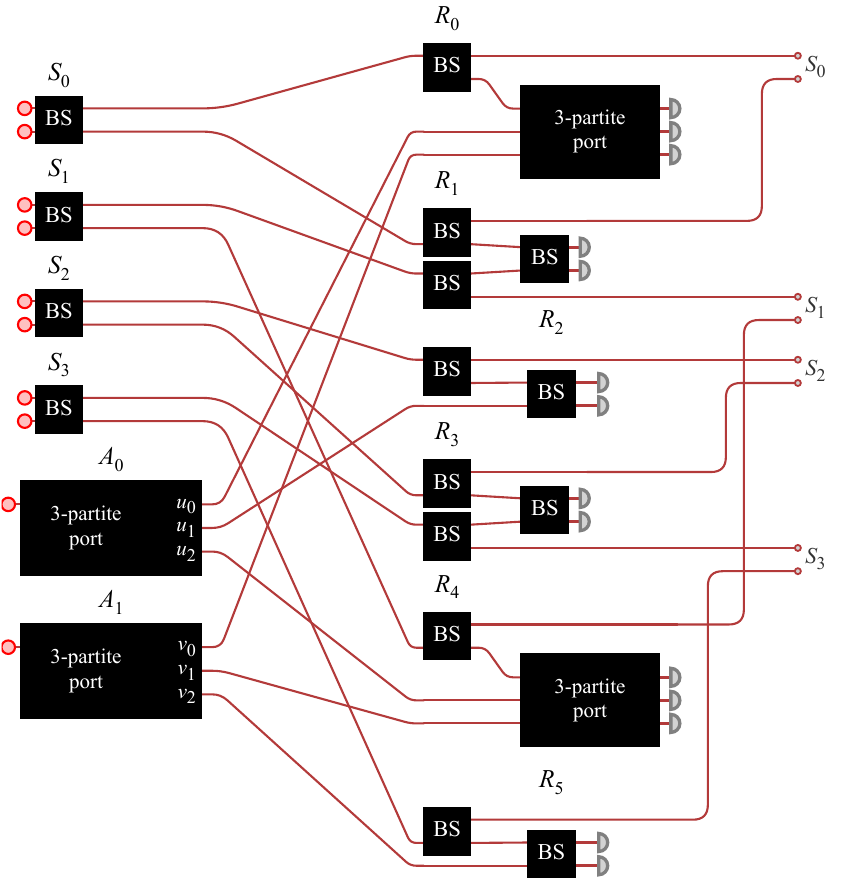}
        \caption{Linear optical circuit for generating the $[[4,1,2]]$ error correcting code. The labels $u_k$ and $v_k$ ($k\in \{0,1,2\}$) on the ancilla outputs indicate which amplitude of Eq.~\eqref{eq:qec_amplitude_vectors} each rail carries.}
        \label{fig:N_4_qec_circuit}
\end{figure}

\paragraph*{[[4,1,2]] code.---} This is a stabilizer code that encodes a logical qubit as
    \begin{align}
        &|0_L\> = \frac{1}{\sqrt{2}}(|0000\> + |1111\>),\\
        &|1_L\> = \frac{1}{\sqrt{2}}(|0011\>+ |1100\>),
    \end{align} which has 3 stabilizer generators
    \begin{align}
    \< Z_0Z_1, Z_2Z_3, X_0X_1X_2X_3\>.    
    \end{align}
    This is called the $N=4$ amplitude-damping code~\cite{leung1997approximate}. It detects single-qubit Pauli errors and protects against single-qubit losses. This loss-tolerant feature is particularly useful for linear optical quantum computing, because it can substantially reduce logical errors in loss-dominated photonic systems despite its simple structure.  

Our graph search finds a heralded generation scheme for the code in the $(4,2)$ repository.
Our graph solutions to generate 
\begin{align}
    |\text{ADC}\> = \a|0_L\> + \b|1_L\>  
\end{align} ($|\a|^2+ |\b|^2=1$) with arbitrary $\a$ and $\b$ are displayed in Fig.~\ref{fig:qec} (a). 
The circuit for the code is presented in Fig.~\ref{fig:N_4_qec_circuit}. For any $(\a,\b)$ with $\a\b\neq0$, the optimal output amplitudes yield $P_{\text{succ}} = 1/2^9$.
    
We can consider the general $[[2k,1,2]]$ loss-tolerant stabilizer code ($k\ge 2$) as
    \begin{align}
        &|0_L\> = \frac{1}{\sqrt{2}}(|0\>^{\otimes 2k } + |1\>^{\otimes 2k}),\\
        &|1_L\> = \frac{1}{\sqrt{2}}(|0\>^{\otimes k}|1\>^{\otimes k } + |1\>^{\otimes k}|0\>^{\otimes k }). 
    \end{align}   
    with $2k-1$ stabilizer generators
\begin{align}
    &\<Z_0Z_1, Z_1Z_2,\cdots, Z_{k-2}Z_{k-1}, \nn \\
    &~~~Z_kZ_{k+1},\cdots, Z_{2k-2}Z_{2k-1}, X_0\cdots X_{2k-1} \>.
\end{align}
It detects single-qubit Pauli errors, corrects single-photon losses, and corrects $(k-1)$-photon losses confined to either the first or the last block of $k$ physical qubits.
We found a heralded generation scheme for [[6,1,2]] in the $(6,2)$ repository as in Fig.~\ref{fig:qec} (b) which achieves $P_{\text{succ}} = 1/2^{13}$ for $\a\b\neq0$. The corresponding circuit is in Appendix~\ref{appendix:qec6} with a detailed analysis.


Although our graph enumeration and search are performed for specific system sizes, the resulting graph solutions in Fig.~\ref{fig:qec} reveal a recurring structural pattern.  Based on this, we derive a general scheme for a general $[[2k,1,2]]$ code (Fig.~\ref{fig:qec} (c)) with $P_{\text{succ}} = 1/(2\cdot 4^{2k})$ for $\a\b\neq0$. One can see that the graph actually generates the $[[2k,1,2]]$ code by verifying the perfect matchings of the graph (Appendix~\ref{appendix:LQG}).
\emph{This demonstrates that the numerical graph repository can provide insight into general graph constructions beyond the finite cases considered in the search.}

\paragraph*{$N=3$ amplitude-damping code.---} Our search also identifies 11 heralded schemes for preparing the logical codewords of the recently proposed three-qubit amplitude-damping code~\cite{dutta2026smallest},
\begin{align}\label{eq:3adc_codewords}
  &|0_L\> = \frac{1}{\sqrt{3}}(|001\> + |010\> + |100\>) \nn \\
  &|1_L\> = |111\>,
\end{align}
in the $(3,3)$ repository.
We present a solution with the best success probability in Fig.~\ref{fig:qec} (d). This graph generates the real-amplitude family
\begin{align}
\a|0_L\> + \b|1_L\>~~~(|\a|^2+|\b|^2 =1)
\end{align} with the success probability
\begin{align}\label{eq:adc3_closed_form}
P_{\text{succ}}(\chi) =
 \frac{(3+\chi^2)\left(1+\sqrt{4\chi+5}\right)^4}
 {3072\,(1+\chi)\left(2+\chi+\sqrt{4\chi+5}\right)^3}
\end{align} where $\chi \equiv \sqrt{3}\,|\b/\a|$, the ratio of the $|111\>$ and $|001\>$ amplitude magnitudes.

It reaches its minimum at $\chi\approx1.41$, where
$P_{\text{succ}}\approx0.75\times10^{-3}$, and increases toward the two
endpoints, approaching $4.23\times10^{-3}$ as $\b\to 0$ and $1/64$ as
$\a\to 0$. 
The corresponding circuit with a more detailed analysis is  in Appendix~\ref{appendix:N3_ADC}.

\subsection{General three-qubit states}

\begin{figure*}[t]
    \centering
        \paperpanel{0.28\textwidth}{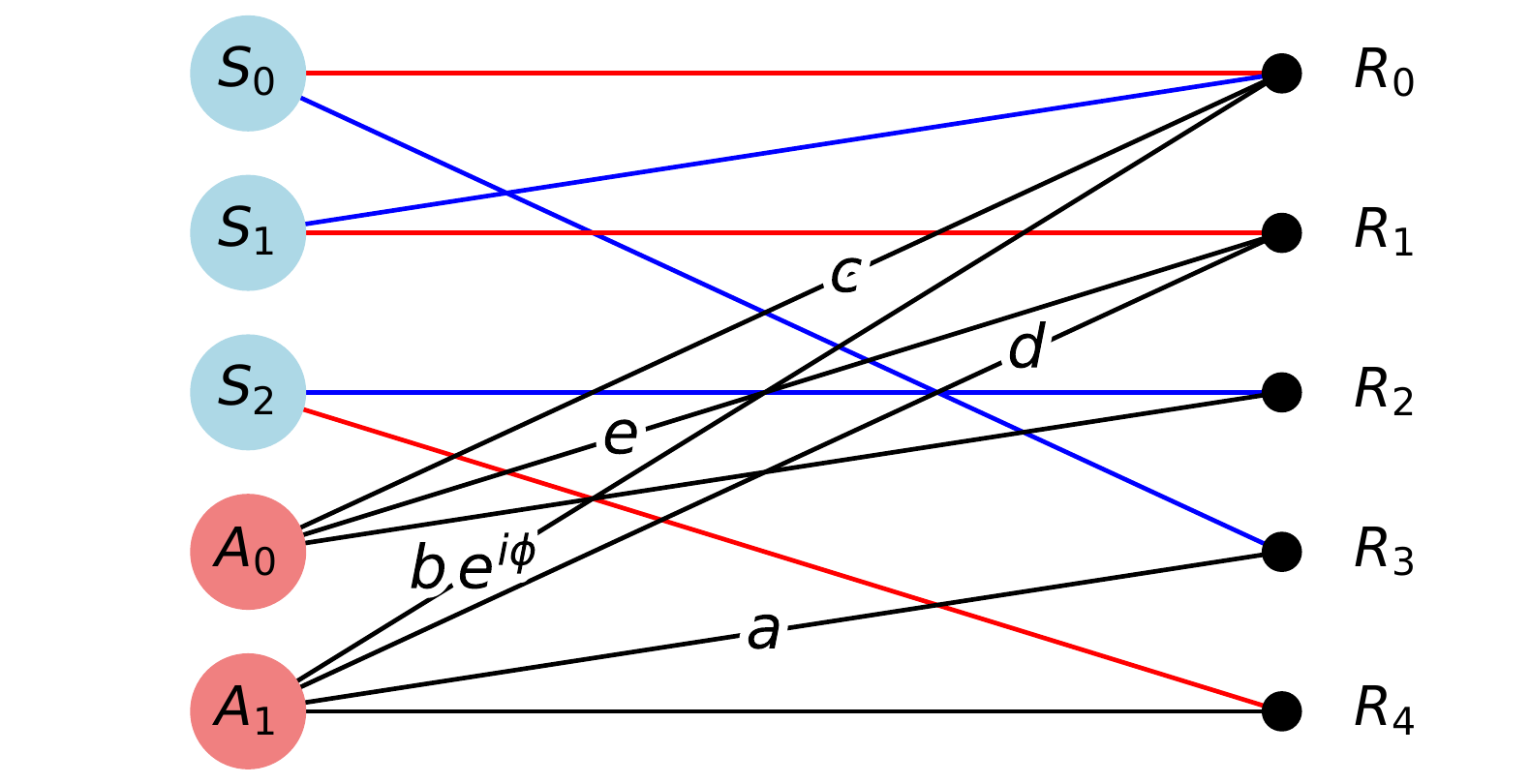}{a}\hfill
        \paperpanel{0.28\textwidth}{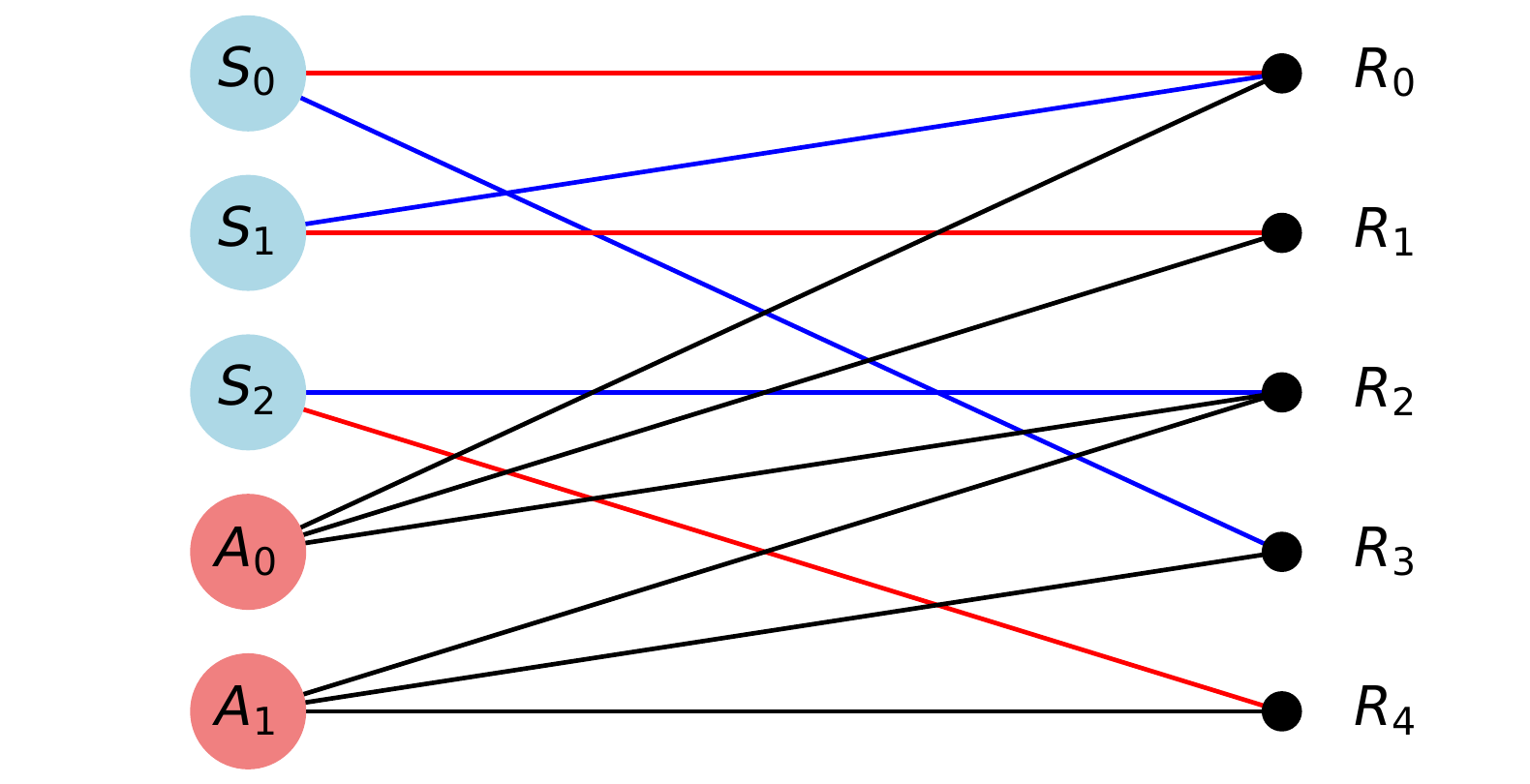}{b}\hfill
        \paperpanel{0.28\textwidth}{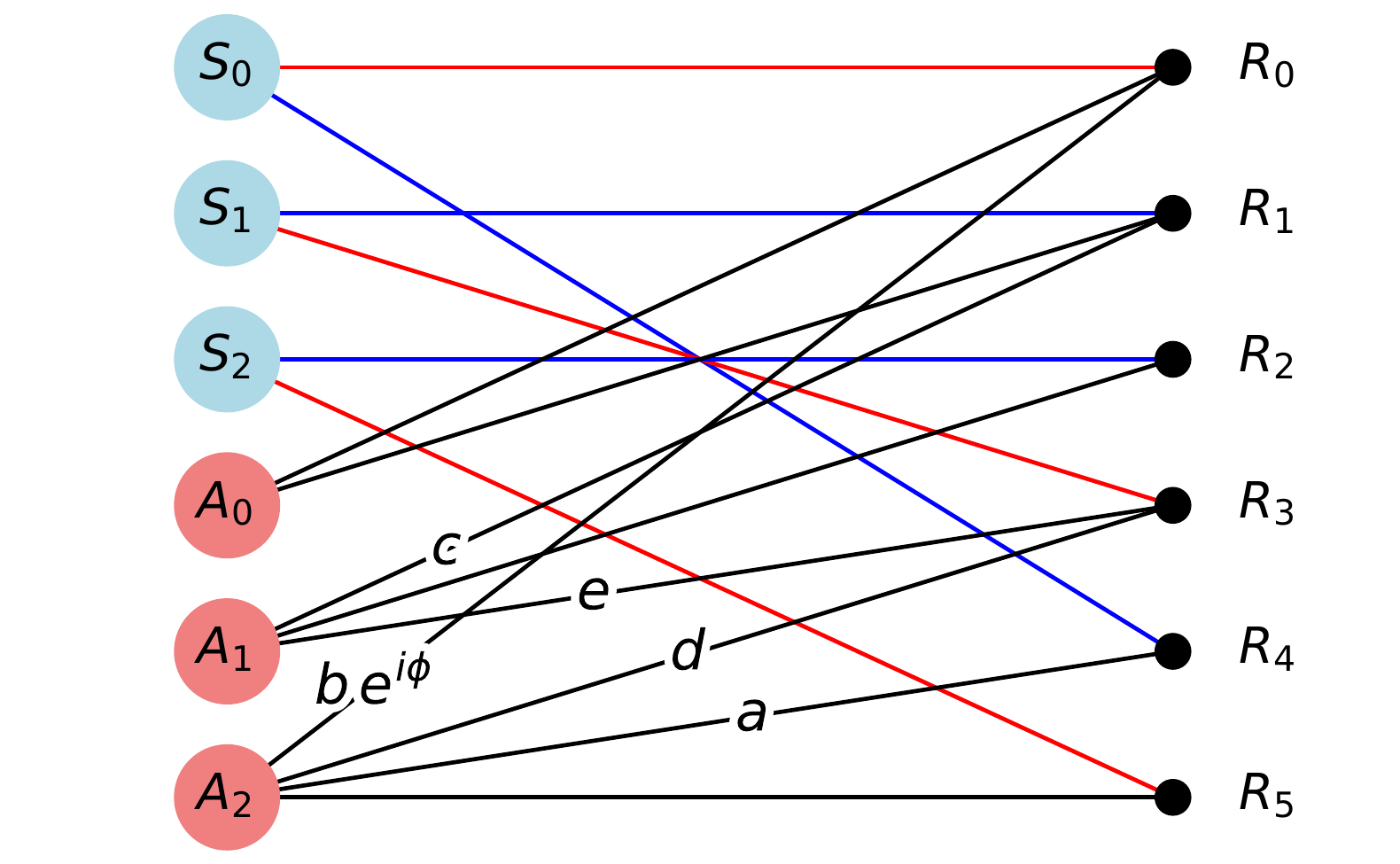}{c}
        \caption{EPM graph solutions for general three-qubit states. (a) and (c) generate the arbitrary tripartite state~\eqref{eq:arbit_tripartite}, found in the $(3,2)$ and $(3,3)$ repositories, respectively, whereas (b) generates Type 5 states. Only the free-parameter edges are labeled by the target coefficients of Eq.~\eqref{eq:arbit_tripartite}.
        }
        \label{fig:type5}
\end{figure*}

In the $N=3$ qubit space, the GHZ and W states are two of the best-known examples of GME states. More general three-qubit states can be represented in the generalized tripartite Schmidt-decomposed form~\cite{acin2000generalized} as
\begin{align}\label{eq:arbit_tripartite}
a|000\> + b\,e^{i\phi}|100\> + c|101\> + d|110\> + e|111\>
\end{align} ($a,\dots,e$ and $\phi$ are non-negative real numbers and satisfy $a^2+b^2+c^2+d^2+e^2 =1$).
Only a postselected linear optical scheme for the arbitrary tripartite states was suggested~\cite{blasiak2022arbitrary}. 

The $(3,2)$ repository includes a graph for the heralded generation of Eq.~\eqref{eq:arbit_tripartite} (Fig.~\ref{fig:type5} (a)). See Appendix~\ref{appendix:type5} for the corresponding linear optical circuit and its detailed analysis.
Our scheme generates the state with the same $(a,b,c,d,e)$ but 12 different values of $\phi$ with the total success probability 
\begin{align}\label{eq:type5_closed_form}
    P^{(3,2)}_{\text{succ}} = \frac{1}{64\left(\sqrt{a^2+b^2+d^2}+\sqrt{c^2+e^2}\right)^2}.
\end{align}
This ranges from \(1/128\) when $a^2+b^2+d^2=c^2+e^2=1/2,$ toward $1/64$ as either sum approaches zero. 
When $(b,c,d,e)$ are all nonzero, the success probability for a specific value of $\phi$ is given by $P_{succ}^{(3,2)}/12$. When any of the variables is zero, the success probability for the resulting state is equal to $P_{succ}^{(3,2)}$. See Appendix~\ref{appendix:type5} for the corresponding circuit and analysis.

Using this scheme, we can generate any tripartite entangled states by setting the amplitudes properly. For example, we can generate $N=3$ magic state with $P_{\text{succ}}=1/128$, the same value as that of the circuit in Fig.~\ref{fig:magic3_circuit}. As another example, by setting $a=b=c=d=e=\frac{1}{\sqrt{5}}$ and $\phi =0$, we generate an $N=3$ Type 5 state~\cite{acin2000generalized}
\begin{align}\label{eq:type_5}
   \frac{1}{\sqrt{5}}\big(|000\> + |100\> + |101\> + |110\> + |111\>\big),
\end{align} which is not LU-equivalent to either the canonical GHZ state or the W state. Previously,
Ref.~\cite{chin2024heralded} suggested a scheme to generate the state 
using 9 photons with success probability $P_{\text{succ}} = \frac{5}{2^73^2} \approx 4.34\times 10^{-3}$.
Our graph solution in Fig.~\ref{fig:type5} (a) can generate the same state with 8 photons and
$P_{\text{succ}} = 5(5-2\sqrt6)/768 \approx 6.58\times 10^{-4}$. On the other hand, we can also find another graph  as in Fig.~\ref{fig:type5} (b) that generates the state
with 8 photons and $P_{\text{succ}} = 5/(2^6 3^2) \approx 8.68\times 10^{-3}$,
exactly twice the probability of Ref.~\cite{chin2024heralded}
(Appendix~\ref{appendix:type5}).

The $(3,3)$ repository also contains 16 graphs that can generate the same state. We present one of them in Fig.~\ref{fig:type5} (c), which generates states with three different values of $\phi$. The total success probability is given by
\begin{align}\label{eq:w2_general_success}
&P_{\text{succ}}^{(3,3)}=\frac{z}{64(1+z)\big(\sqrt{c^2+e^2}+\sqrt{z(a^2+b^2) +d^2}\big)^2},
\end{align} where $z$ is a tunable parameter that can be optimized to maximize the success probability. 
We can show that this scheme can generate  many tripartite states with higher success probability than the scheme corresponding to Fig.~\ref{fig:type5} (a), at the cost of one additional photon (Appendix~\ref{appendix:type5}).



\subsection{$N=4$ cluster and caterpillar graph states}

Cluster states are fundamental resources for quantum information processing. They are the backbone of MBQC and are also widely used in quantum communication, quantum networks, and quantum error correction~\cite{raussendorf2001one,azuma2015all}. 
The $N=4$ cluster state
\begin{align}\label{eq:N=4_cluster}
 \frac{1}{2}(|0000\> +|0011\> +|1100\> -|1111\>)    
\end{align}
is the smallest nontrivial cluster state that serves as a fundamental building block for larger cluster-state architectures. It is well known that this can be generated by fusing entangled resource states such as Bell states and $N=3$ GHZ states~\cite{browne2005resource,varnava2006loss}. A monolithic generation from single-photon initial states was also proposed in Ref.~\cite{chin2026efficient}.

We found a scheme in the (4,2) repository that generates the state using 10 photons.
Its heralding outcomes generate three phase-distinct weighted cluster states~\cite{hartmann2007weighted,plato2008random}, differing by the relative phase of $|1111\>$,
with $P_{\text{succ}} = 1/512\approx 1.95\times 10^{-3}$ (Fig.~\ref{fig:cluster4}, see Appendix~\ref{appendix:cluster4} for the corresponding linear optical circuit).
One third of the successful outcomes generate the standard $N=4$ cluster state~\eqref{eq:N=4_cluster}, corresponding to $P_{\text{succ}}=1/1536$.

Our scheme uses the same number of photons as the construction based on fusing an $N=3$ GHZ state with a Bell state~\cite{chin2026efficient}. For the exact generation of the standard cluster state, its success probability is one third of that fusion-based construction. On the other hand, the monolithic architecture directly generates a family of weighted cluster states within a single optical network, without separate preparation, synchronization, and fusion of multiple entangled resource states.
A recent work~\cite{hartnett2026automated} proposed an optimized scheme using 9 photons with a success probability of $2.05\times10^{-3}$, although it requires a more complicated circuit containing 45 beam splitters  (see Appendix~\ref{appendix:cluster4} for a comparison). 

\begin{figure}[t]
    \centering
    \begin{minipage}[b]{0.45\linewidth}
        \centering
        \includegraphics[viewport=64.64 321.67 239.67 497.27,clip,width=.24\linewidth]{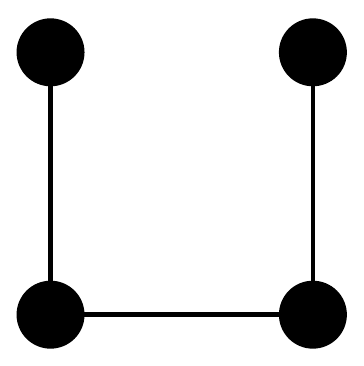}\par\vspace{0.4em}
        \includegraphics[viewport=447.49 174.49 1185.99 643.79,clip,width=\linewidth]{figure/cluster4.pdf}\par
        \normalfont(a)
    \end{minipage}\hspace{0.04\textwidth}%
    \begin{minipage}[b]{0.45\linewidth}
        \centering
        \includegraphics[viewport=7.19 403.65 255.04 560.33,clip,width=.34\linewidth]{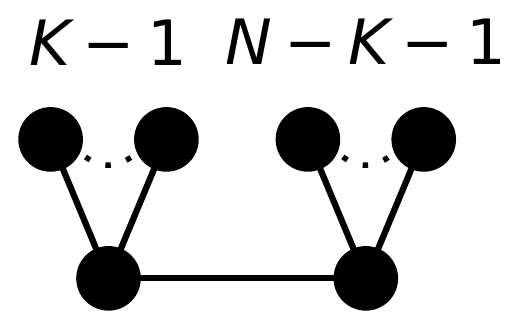}\par\vspace{0.4em}
        \includegraphics[viewport=380.56 19.18 1112.11 830.50,clip,width=\linewidth]{figure/caterpillar1.pdf}\par
        \normalfont(b)
    \end{minipage}
    \caption{Graph representations of (a) the $N=4$ linear cluster-state solution found in the $(4,2)$ repository and (b) the arbitrary $N$-partite length-1 caterpillar graph-state construction inferred from the pattern in (a). In each panel, the target graph state is shown above and the corresponding EPM bigraph below.}
    \label{fig:cluster4}
    \label{fig:caterpillar}
\end{figure}

The structure in Fig.~\ref{fig:cluster4}(a) can be generalized to generate arbitrary $N$-partite length-1 caterpillar graph states
 \begin{align}
    &\frac{1}{2}\big(|\underbrace{0\cdots 0}_{K}\underbrace{0\cdots 0}_{N-K}\> +|\underbrace{0\cdots 0}_{K}\underbrace{1\cdots 1}_{N-K}\> \nn \\
    &~~~~~~+ |\underbrace{1\cdots 1}_{K}\underbrace{0\cdots 0}_{N-K}\> -  |\underbrace{1\cdots 1}_{K}\underbrace{1\cdots 1}_{N-K}\>\big)   
 \end{align}
 as in Fig.~\ref{fig:caterpillar}(b), which can be verified by considering all the perfect matchings of the graph (Appendix~\ref{appendix:LQG}). Compared to the known monolithic scheme in Ref.~\cite{chin2026efficient}, our scheme requires one fewer photon.
 


\section{Discussion}\label{sec:discussions}

In this work, we have introduced a systematic algorithmic approach based on the LQG picture for designing heralded linear optical circuits that generate GME states. The schemes presented in this work were selected to illustrate the breadth of the proposed framework across several types of multipartite resource states.
A crucial advantage of the framework is that the numerical graph repository serves not only as a source of individual solutions but also as a means of discovering general graph constructions. 
Moreover, once a repository is constructed, it can be preserved and reused to search for other  interesting target states that are not considered here.

It is worth comparing our method with previous approaches to the automated design of optical experiments. First, general-purpose optical-design tools such as THESEUS, Klaus, and PyTheus~\cite{krenn2021conceptual,cervera2022design,ruiz2023digital} explore large experimental search spaces using stochastic or graph-theoretic heuristics. The search is carried out directly over optical elements or their graph representations.  This makes the search for heralded schemes non-trivial, as the constraints associated with ancillary photons, heralding measurements, and quantum interference are treated directly. Second, Ref.~\cite{hartnett2026automated} focuses on numerical optimization of individual heralded circuits for small stabilizer graph states. Although effective for obtaining specific solutions, the method and results are restricted to small-sized stabilizer states and lack general patterns or scalable design principles. In contrast, our algorithmic method reduces the experiment space in the LQG picture for designing heralded schemes and is useful for both stabilizer and non-stabilizer states. It also systematically verifies circuit structures based on the patterns of their graph representations for general $N$-partite states.

Currently, our graph repositories are limited to systems with relatively few ancilla modes due to their computational cost. Since the ancillae provide additional structural degrees of freedom for generating diverse GME states, we can try to find more efficient algorithms to increase the ancilla size. This can be achieved by finding better algorithmic strategies, or imposing stronger heuristic restrictions on the graphs for the reduction of  the search space.

Additionally, while EPM graphs capture a useful restricted class of heralded linear-optical schemes, our current graph formalism is not sufficiently general to represent all such circuits. A more general framework will therefore be required for a more comprehensive design of heralded linear-optical schemes.

Finally, in the present framework, circuit design is achieved indirectly through graph enumeration, graph search, and graph-to-circuit translation. Whether a fully constructive design methodology exists for arbitrary heralded state generation remains an open question. Nevertheless, developing such direct design methodologies remains an important direction for future research.

\section*{Data Availability}
The data and code that support this research are available from SC upon reasonable request.

\section*{Acknowledgements}

The authors are grateful to Peizhe Li for fruitful discussions.
SC was supported by National Research Foundation of Korea (NRF, RS-2023-00245747).
JH acknowledges support by the National Research Foundation of Korea (NRF, RS-2023-NR068116, RS-2025-03532992), the Institute for Information \& Communications Technology Promotion (IITP, No. 2019-0-00003), and the Korea Health Industry Development Institute (KHIDI, RS-2025-25456722).
WJM was supported in part by the Japan’s Council for Science, Technology and Innovation (CSTI) under the Cross-ministerial     Strategic Innovation Promotion Program (SIP) for “Promoting the     application of advanced quantum technology platforms to social     issues”(JPJ012367).

\appendix

\section{Review of the linear quantum graph (LQG) picture}\label{appendix:LQG}

A heralded scheme signals the successful preparation of a target state through measurement outcomes on ancillary modes without measuring the prepared state, which therefore stays  available for further use~\cite{forbes2025heralded}.
These heralding detectors after linear transformations can be abstracted as \emph{boson subtraction operators}. 

In our setup~\cite{chin2024shortcut}, 
each particle has a spatial state and a qubit state, whose creation and annihilation operators are denoted as $\ha_{j,s}^\dagger$ and $\ha_{j,s}$ ($j\in \{0,1,\cdots, N+M-1\}$, $s \in \{+,-\}$) respectively. Then the initial state is set to contain two photons in distinct internal states $|+\>$ and $|-\>$ in the $N$ main systems and one photon with a qubit state $|+\>$ in the $M$ ancillary systems: 
	\begin{align}\label{eq:initial}
		|\Psi_\text{Initial}\> \equiv  \prod_{j=0}^{N-1}(\ha^\dagger_{j,+}\ha^\dagger_{j,-})\prod_{j=N}^{N+M-1}\ha^\dagger_{j,+}|vac\>. 
	\end{align} 
We can generate $N$-partite entangled states by applying an $N+M$ spatially overlapped subtraction operators
\begin{align}\label{eq:subtraction}
		&\prod_{l=0}^{N+M-1}\sum_{j=0}^{N+M-1}(k^{(l)}_{j,+}\ha_{j,+} + k^{(l)}_{j,-}\ha_{j,-} ) \nn \\
		&\qquad\equiv \prod_{l=0}^{N+M-1}\hat{A}^{(l)} \equiv \hat{A}_{N+M}  \nn \\
       &~~~~~~~~~~~~\quad (k^{(l)}_{j,s} \in \mathbb{C}~\textrm{and}~  \sum_{j,s}|k^{(l)}_{j,s}|^2 =1).
	\end{align}  

The operator $\hat{A}_{N+M}$ must satisfy the \emph{no-bunching condition}~\cite{chin2024shortcut}, meaning that the final state
 $|\Psi_\text{final}\>  =  \hat{A}_{N+M}|\Psi_\text{Initial}\>$ 
is supported on states with exactly one particle occupying each mode. In the context of heralded state generation, the particle remaining in each mode encodes the qubit, while the subtracted particle serves as the heralding signal. Then the above operation corresponds to a heralded scheme with $2N+M$ photons and $N+M$ detectors. Our goal is to find a suitable operator  $\hat{A}_{N+M}$ that generates the expected multipartite entangled state. 

We can simplify the circuit design problem by mapping the operator into graph elements in the LQG picture, which is summarized in Table~\ref{tab:dictionary}.	In this picture, the no-bunching condition is equivalent to the condition that the subtraction operators are arranged such that the supports of final state $|\Psi_\text{final}\>$ are mapped to the perfect matchings (PMs) of the graph. In general, however, it is not straightforward to assign suitable edge weights that simultaneously satisfy this condition and generate GME.

\begin{figure}[t]
    \centering
    \includegraphics[width=0.9\linewidth]{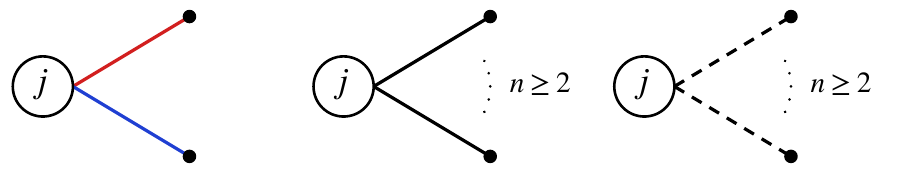}
    \caption{EPM bigraph configurations~\cite{chin2024shortcut}.
Solid black, dashed black, red, and blue edges denote
the internal states $|+\>$, $|-\>$, $|0\>$, and $|1\>$,
respectively.}
    \label{fig:epm_bigraph}
\end{figure}

To overcome this difficulty, we can impose two restrictions on the graphs that reduce the search space significantly. First, we can define a convenient class of graphs that automatically satisfy the no-bunching condition, which is called \emph{effective perfect matching graphs} (EPM graphs)~\cite{chin2024shortcut}.  The edges of EPM graphs are attached to the circles as in Fig.~\ref{fig:epm_bigraph}. One can check that if a graph is an EPM graph, then all the supports of the final state correspond to the PMs of the graph~\cite{chin2024shortcut}. Second, all the graph solutions that generate GME must be \emph{strongly connected} in their directed graph representation~\cite{chin2021graph,chin2026efficient} (see Definition~\ref{def:epm_digraph} below). In the corresponding directed graphs, perfect matchings of the graphs are represented as disjoint cycle covers. 
 Consequently, our task in Stage 1 is to enumerate strongly connected EPM graphs.  
 
\begin{table}[t]
\centering 
			\begin{tabular}{|l|l|l|}
				\hline
				\textbf{Subtraction operators} & \textbf{Bipartite graphs}   \\
				\textbf{in photonic systems} & ~~~~~in \textbf{$G =(U\cup V, E)$}\\
				\hline\hline 
				Spatial modes $j$ & Labeled circles (\encircle{$j$}) $\in$ $U$ \\ \hline 
				$\hat{A}^{(l)}$  
    &     Unlabeled dots ($\bullet$) $\in$  $V$  \\ \hline 
				Spatial distributions of $\hat{A}^{(l)}$  & Undirected edges  $\in$ $E$  \\ \hline 
				Probability amplitude $\a_j^{(l)} $ & Edge weight $\a_j^{(l)} $ \\ \hline 
				Internal state $\p_j^{(l)} $ & Edge weight $\p_j^{(l)}$  \\
				\hline 
			\end{tabular}
			\caption{Mapping relations between subtraction operators and bipartite graphs in the LQG picture.
            In this work we only consider four qubit states $\{|+\>,|-\>, |0\>,|1\>\}$  $\big(|0\> = \frac{1}{\sqrt{2}}(|+\> + |-\>)$, $|1\> = \frac{1}{\sqrt{2}}(|+\> -|-\>) \big)$, hence the edge weight for the qubit states are replaced with edge colors \{Solid Black, Dashed Black, Red, Blue\} for simplicity. 
            Consequently, whenever we refer to edge weights, we mean the edge weights associated with the probability amplitudes.}
            			\label{tab:dictionary}
		\end{table}

\begin{figure*}[t]
    \centering
    \begin{minipage}[t]{.85\textwidth}
        \centering
        \setlength{\parskip}{0pt}%
        \paperpanel{.5\linewidth}{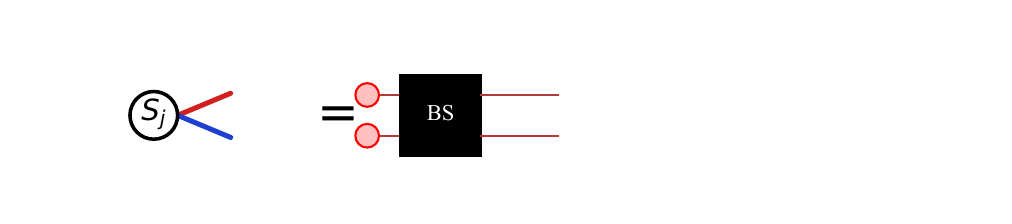}{a}%
        \paperpanel{.5\linewidth}{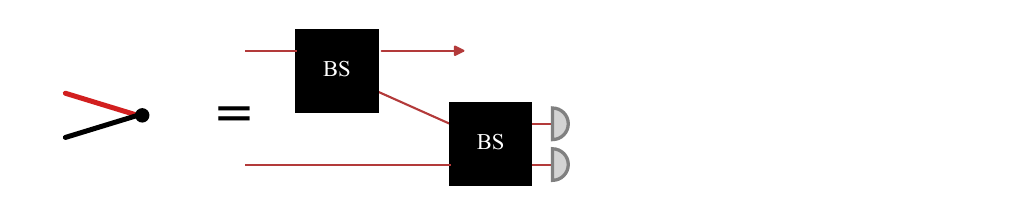}{d}
        \par\vspace{2pt}%
        \paperpanel{.5\linewidth}{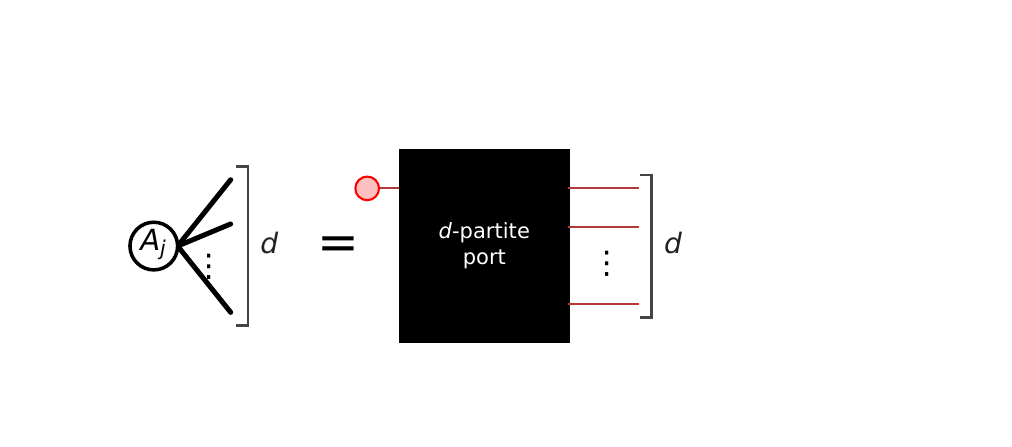}{b}%
        \paperpanel{.5\linewidth}{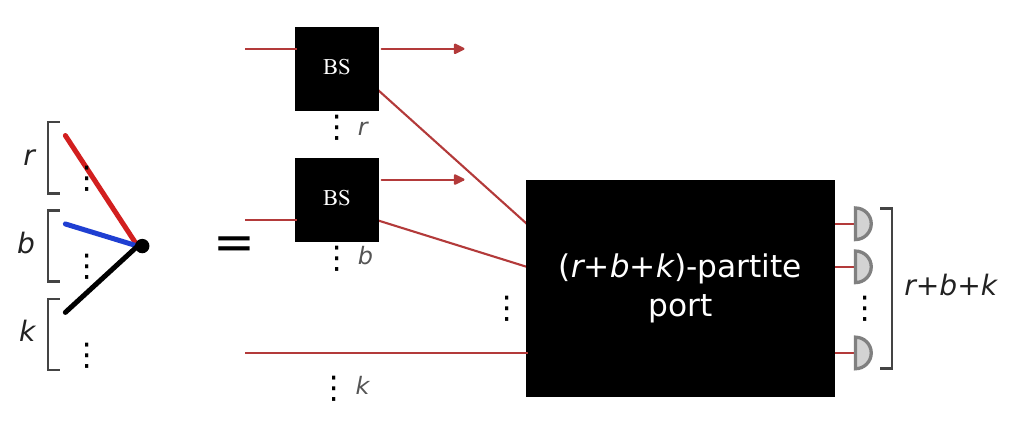}{e}
        \par\vspace{2pt}%
        \paperpanel{.5\linewidth}{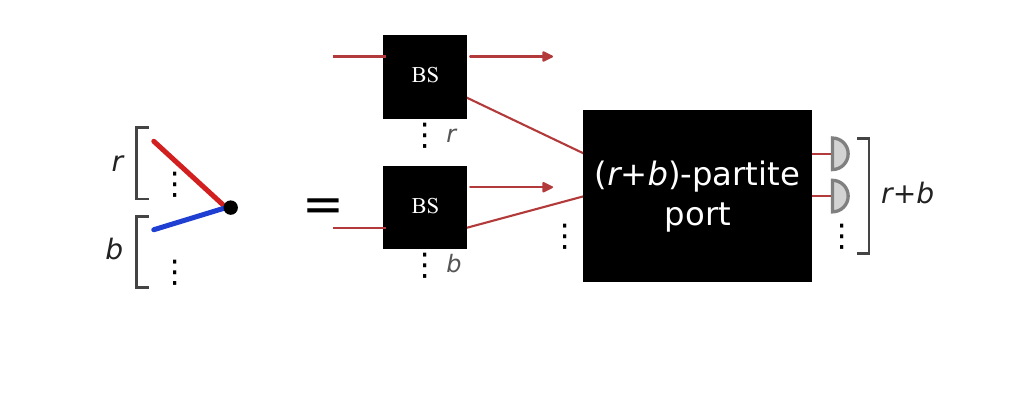}{c}%
        \paperpanel{.5\linewidth}{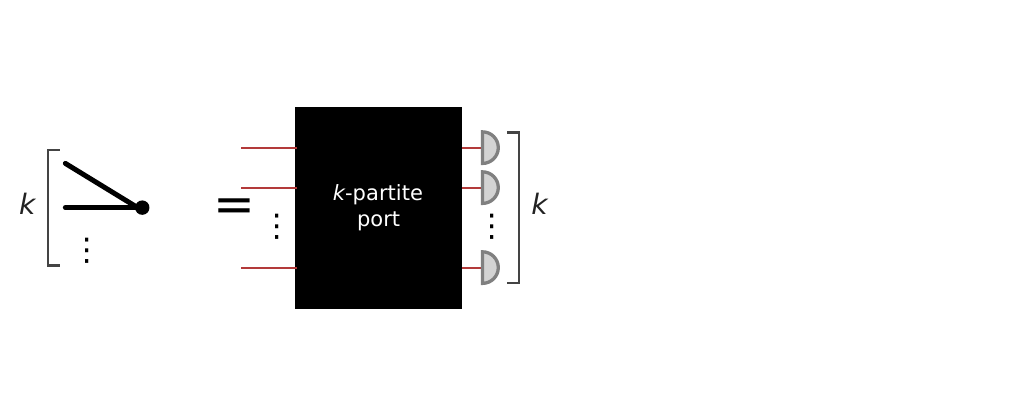}{f}
    \end{minipage}
    \caption{Translation rules from EPM bigraph elements to dual-rail linear-optical elements, adapted from Refs.~\cite{chin2024heralded,kang2026heralded}. Red, blue, and black edges carry the internal states $|0\>$, $|1\>$, and the ancillary photons, respectively; red circles denote single-photon sources, half-disks denote heralding detectors, unterminated arrows denote undetected rails that form the system-qubit outputs, and vertical dots denote repetition. (a)~A system node $S_i$ becomes two sources and a balanced beam splitter (BS), whose upper (lower) output rail follows the red (blue) edge. (b)~An ancilla node $A_j$ of degree $d$ becomes one source and an asymmetric $d$-partite multiport; for $d=2$ the port is a BS with vacuum on the unused input. (c)--(f)~At a photon subtraction node (dot) with $r$ red, $b$ blue, and $k$ black incident edges, each colored edge passes through its own BS; the inner BS outputs and the rails of the black edges interfere in a symmetric $(r{+}b{+}k)$-partite multiport, all of whose outputs are detected. The panels show (c)~colored edges only ($k=0$), (d)~one colored and one black edge, (e)~the general mixed case ($r{+}b\ge 1$, $k\ge 1$, $r{+}b{+}k\ge 3$), and (f)~black edges only ($k\ge 2$). Symmetric multiports implement the discrete Fourier transform; $\mathrm{DFT}_2$ is the balanced BS. }
    \label{fig:translation_rules}
\end{figure*}

Once we find a graph solution for a target state, we need to decode the graph as linear optical components and detectors for an actual circuit design. A set of translation rules from graphs to linear optical circuits was proposed in Ref.~\cite{chin2024heralded} and is summarized in Fig.~\ref{fig:translation_rules} for dual-rail encoding. 

\section{Methods}\label{appendix:methods}

The computation consists of three stages of search and analysis. Throughout, $N$ denotes the number of system (qubit) nodes and $M$ the number of ancilla nodes.
Pseudocode for the routines of this section is provided in the Supplemental Material.

\subsection{Stage 1: Graph enumeration}\label{appendix:stage1}

\paragraph*{Generation.}
Under the conventions of Restrictions~1 and~2, we generate all labeled
graphs satisfying Definition~\ref{def:epm} by direct construction. Each
system node $S_i$ is connected to its fixed partner $R_i$ and to one
additional photon-subtraction node chosen from the remaining $N+M-1$.
Each ancilla node $A_j$ is connected to a subset of at least two of the
$N+M$ photon-subtraction nodes.

Before the minimum-degree filter of Restriction~3, these choices give
the raw candidate count
\begin{equation}
    (N+M-1)^N\bigl(2^{N+M}-1-(N+M)\bigr)^M .
\label{eq:general_ancilla_choice_count}
\end{equation}

\paragraph*{Restriction 1: Fixed partner convention.}
Fixing the partner edge $S_iR_i$ removes most of the relabeling freedom of the photon-subtraction nodes; the residual relabeling duplicates are removed by the canonical-signature reduction below.

\paragraph*{Restriction 2: Red/blue label assignment.}
Assigning the red (blue) label to the lower-indexed (higher-indexed) photon-subtraction neighbor of each $S_i$ selects one representative among the $2^N$ computational-basis label assignments. The discarded label freedom is a local Pauli-$X$ operation on the corresponding qubit and is restored explicitly by the $P\times X$ search of Stage~2.

\paragraph*{Restriction 3: Minimum node degrees.}
We impose $\deg(R_k)\ge 2$ on every photon-subtraction node; the ancilla nodes satisfy $\deg(A_j)\ge 2$ by construction, and the system nodes have degree two by Definition~\ref{def:epm}. A photon-subtraction node of degree zero cannot be covered by any perfect matching, so the graph generates no state. A photon-subtraction node $R_k$ of degree one forces its unique edge into every perfect matching. If that edge is the red or blue edge of a system node $S_i$, the bit of qubit $i$ is fixed in every perfect matching, so the generated state is a product state on that qubit and is not GME. If it is the black edge of an ancilla node $A_j$, the remaining edges of $A_j$ appear in no perfect matching; removing $A_j$, $R_k$, and those edges leaves the generated state unchanged and yields an EPM bigraph with one fewer ancilla node, to which the same argument applies. Excluding photon-subtraction nodes of degree less than two therefore loses no GME state.

\paragraph*{Canonical-signature reduction.}
Each generated graph is canonically relabeled, keeping the three node
types $S$, $A$, and $R$ distinct and disregarding the red/blue edge labels.
The sorted edge list of the canonical form is stored as its canonical
signature. Two graphs have equal canonical signatures if and only if
they are isomorphic under a type-preserving relabeling, up to the label
assignment of Restriction~2. We retain one representative per canonical
signature. Together with the minimum-degree filter of Restriction~3,
this reduction gives the non-trivial canonical graphs listed in
Table~\ref{tab:enumeration}.

\begin{definition}[EPM Directed Graph]
\label{def:epm_digraph}
Let $Q_0,\ldots,Q_{N+M-1}$ denote the ordered nodes $S_0,\ldots,S_{N-1},A_0,\ldots,A_{M-1}$, and pair $R_k$ with $Q_k$ by index. Given an EPM bigraph $G$, the \emph{EPM directed graph} $\vec{G}$ on $S\cup A$ contains an edge $Q_k\to Q_i$ whenever $Q_i$ is adjacent to $R_k$ in $G$~\cite{chin2024shortcut}.
\end{definition}

\paragraph*{Restriction 4: Strong-connectivity condition.}
We retain only graphs whose $\vec{G}$ is strongly connected, the graph-theoretic necessary condition for GME generation (Appendix~\ref{appendix:LQG}).

\begin{definition}[Perfect Matching and Generated State]
\label{def:pm}
A \emph{perfect matching} of an EPM bigraph $G$ is a subset of $E$ such that every node of $G$ is incident to exactly one edge in the subset. The $p$-th perfect matching $M_p$ selects, at each system node $S_i$, either the red or the blue edge and thereby yields the computational basis state $|s_p\>$ with $s_p=b_0 b_1\cdots b_{N-1}$, where
\[
  b_i = \begin{cases}
    0 & \text{red edge of } S_i \in M_p,\\
    1 & \text{blue edge of } S_i \in M_p.
  \end{cases}
\]
The generated state is the superposition of the basis states $|s_p\>$ of all perfect matchings, each weighted by the product of the edge weights of $M_p$.
\end{definition}

\paragraph*{Graphs discarded at repository construction.}
When the repository is built (after Restriction~4), graphs with fewer than two perfect matchings cannot produce a superposition and are discarded; graphs with an edge that appears in no perfect matching are discarded as redundant, since deleting the unused edges leaves the generated state unchanged.

\paragraph*{Graph repository.}
For each surviving graph $G$, we store its edge list, perfect matchings,
coefficient vector $C_G$, and spectral signature
(Definition~\ref{def:spectral_signature}). The vector $C_G$ is obtained
by counting the perfect matchings for each computational-basis state
and dividing the resulting integer coefficients by their greatest
common divisor. Entries with equal spectral signatures are grouped.
Each target-state search reuses this stored $(N,M)$ repository without
repeating the enumeration.

\subsection{Stage 2: Target state search}\label{appendix:stage2}

\begin{definition}[$P\times LU$ Equivalence]
\label{def:pxlu}
Two states are \emph{$P\times LU$-equivalent} if $|\psi\rangle=(U_0\otimes\cdots\otimes U_{N-1})\,P_\sigma\,|\phi\rangle$ up to a global phase, for a permutation $\sigma\in\mathrm{Sym}(N)$ of the system modes (qubits), acting as $P_\sigma|b_0\cdots b_{N-1}\rangle=|b_{\sigma(0)}\cdots b_{\sigma(N-1)}\rangle$, and local unitaries $U_i\in SU(2)$. \emph{LU equivalence} is the special case $\sigma=\mathrm{id}$.
\end{definition}

\begin{definition}[Spectral Signature]
\label{def:spectral_signature}
The \emph{spectral signature} of $|\psi\rangle$ collects the eigenvalues of the reduced density matrices over all bipartitions,
\begin{equation}
\Lambda(|\psi\rangle)=\mathrm{sort}\Bigl(\bigsqcup_{\substack{\mathcal{S}\subsetneq[N]\\ 1\le|\mathcal{S}|\le\lfloor N/2\rfloor}}\mathrm{spec}^{+}\bigl(\rho_{\mathcal{S}}\bigr)\Bigr),
\label{eq:spectral_signature}
\end{equation}
where $[N]=\{0,\ldots,N-1\}$, $\rho_{\mathcal{S}}=\mathrm{tr}_{[N]\setminus\mathcal{S}}(|\psi\rangle\langle\psi|)$, and $\mathrm{spec}^{+}$ denotes the positive eigenvalues.
\end{definition}

\paragraph*{Signature invariance and stored signature.}
Local unitaries preserve every spectrum in Eq.~\eqref{eq:spectral_signature}, and a permutation of the qubits only permutes which subset contributes which eigenvalues, so the spectral signature is invariant under the $P\times LU$ equivalence of Definition~\ref{def:pxlu}. For the stored signature, eigenvalues below $10^{-9}$ are discarded and the remaining ones are rounded to six decimal places before sorting.

\paragraph*{Repository groups and GHZ/W filtering.}
From the signature groups of the repository, entries whose signature coincides with that of the ideal GHZ or W state of the same qubit number are removed, since heralded schemes for these states are already known~\cite{chin2024shortcut,chin2024heralded}; the underlying graphs remain in the repository, so these entries can be reconstructed if such targets are considered.

\paragraph*{Spectral-signature matching.}
Given a target state $|\psi_T\>$ with coefficient vector $C_T$, we compute
$\Lambda(|\psi_T\>)$ and select the repository entries with the same
spectral signature as candidates. A matching spectral signature does
not by itself establish $P\times LU$ equivalence; the candidate states
are tested by the searches below. If no group matches, this lookup
returns no candidates.

\begin{definition}[$P\times X$ Transformation]
\label{def:px}
A $P\times X$ transformation is the discrete case of Definition~\ref{def:pxlu} in which every $U_i$ is the identity or the Pauli-$X$ operator: it combines a system-mode permutation $\sigma\in\mathrm{Sym}(N)$ with a bit-flip pattern $f\in\{0,1\}^N$ through $X_f=X^{f_0}\otimes\cdots\otimes X^{f_{N-1}}$, acting as
\[
X_f\,P_\sigma\,|b_0\cdots b_{N-1}\rangle=|b_{\sigma(0)}\oplus f_0,\ldots,b_{\sigma(N-1)}\oplus f_{N-1}\rangle.
\]
\end{definition}

\paragraph*{$P\times X$ search.}
All $N!\,2^N$ images of $C_T$ are precomputed and compared with each candidate for exact equality; the bit-flip patterns of a match restore the system-edge label assignments fixed during generation in Stage~1.

\paragraph*{Numerical $P\times LU$ check.}
If the $P\times X$ search finds no match, we perform the numerical
$P\times LU$ check.
For each system-mode permutation $\sigma$ compatible with the one- and two-qubit reduced-density-matrix spectra of the target, we numerically maximize the fidelity
\begin{equation}
F(\{U_i\};\sigma)=\bigl|\langle\psi_T|(U_0\otimes\cdots\otimes U_{N-1})\,P_\sigma|\psi_G\rangle\bigr|^2
\label{eq:lu_fidelity}
\end{equation}
over single-qubit unitaries $U_i\in SU(2)$, using a three-step optimization: alternating least squares with random restarts, L-BFGS-B refinement, and a differential-evolution fallback. A candidate is accepted when $F\ge 1-10^{-8}$, with the residual and the maximum entry error recomputed on the full coefficient vector. The optimization certifies equivalence in one direction only: $F\approx 1$ confirms a match, while failure of the optimizer does not prove inequivalence.

\paragraph*{Signed targets.}
Stored coefficient vectors are perfect-matching counts and hence non-negative, so a signed target never matches directly. Sign patterns reachable by local unitaries may be found by the numerical check applied to the signed target. If that check finds no match, we turn to the edge weights, the degree of freedom deferred in Stage~1. We run the $P\times X$ search on the absolute-value target and then flip the signs of selected edge weights. Flipping an edge set multiplies each perfect-matching contribution by $(-1)^{q}$, where $q$ counts the flipped edges in that matching; the set is chosen to affect only the basis states that require a negative sign. The recomputed signed vector is kept only when it equals the target exactly, a verification rather than an optimization over sign assignments.

\paragraph*{Parametrized targets.}
For targets with tunable amplitude ratios, we first apply the
$P\times X$ search to the target with all nonzero amplitudes set equal.
The matched graphs generate the same computational-basis terms as the
target, up to the $P\times X$ transformation.

For each edge $e$, we identify the basis states of the perfect matchings
containing it,
\[
\mathrm{affected}(e)=\{s_p:e\in M_p\}.
\]
Basis states sharing a target amplitude form an amplitude group.
Comparing $\mathrm{affected}(e)$ with the amplitude groups identifies
whether an edge affects a single group, several groups, or none.
We identify edges that affect basis states in one group without
affecting other groups, and use their weights to tune the target
amplitudes. These may be edges shared by several basis states in the
group or separate edges for individual basis states.

Matched graphs, together with the transformation that maps their generated state to the target, are passed to Stage~3.

\subsection{Stage 3: Circuit design} \label{appendix:stage3}

\begin{figure*}[t]
    \centering
    \paperpanel{0.44\textwidth}{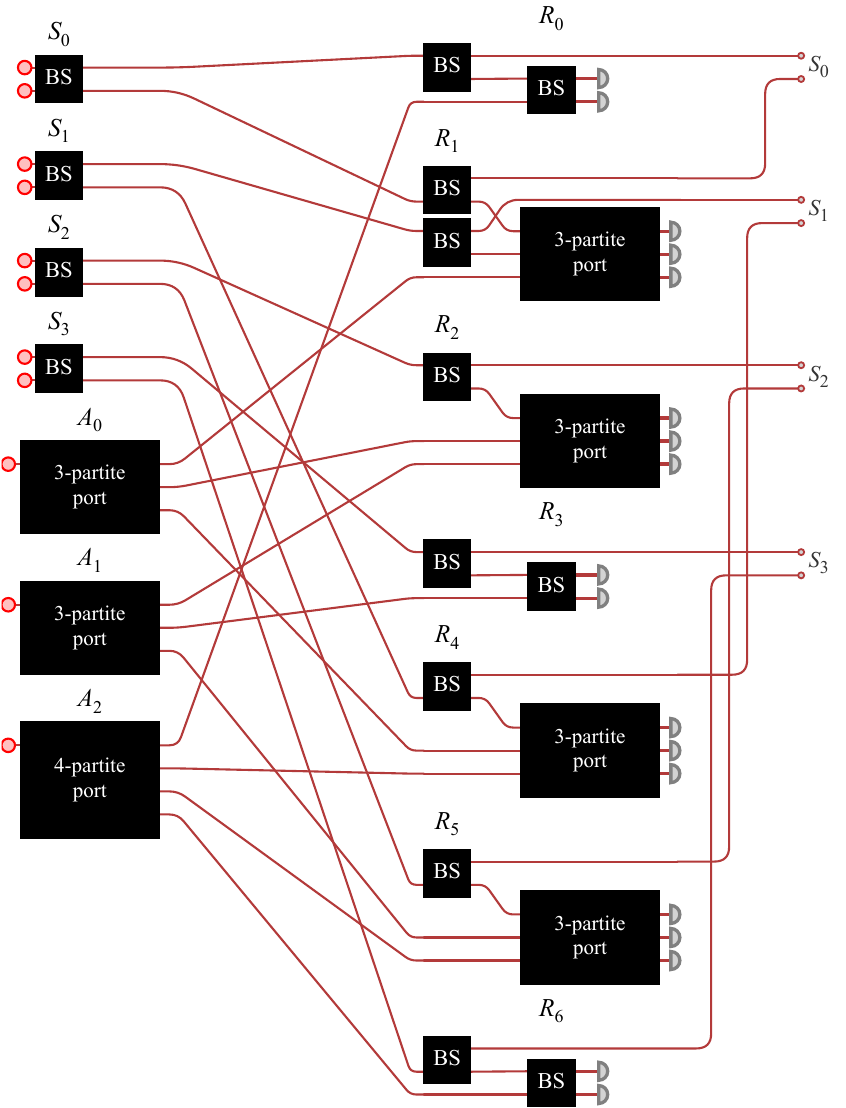}{a}\hfill
    \paperpanel{0.44\textwidth}{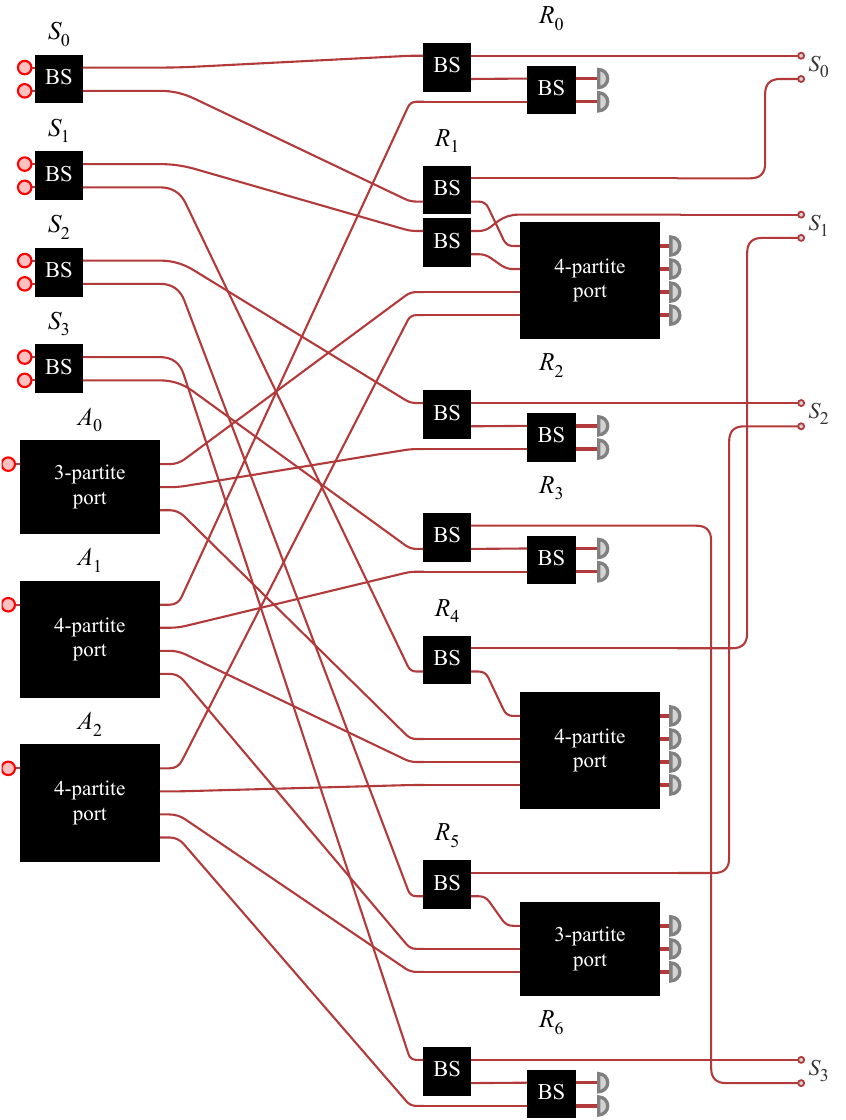}{b}
    \caption{Heralded circuits (a) and (b) generating the $N=4$ hypergraph magic state, translated from the graph solutions in Fig.~\ref{fig:magic_N}(b) and (c), respectively.}
    \label{fig:magic4_circuits}
\end{figure*}


\paragraph*{Dual-rail translation.}
Each node is translated into a block of optical elements, called a panel
(Fig.~\ref{fig:translation_rules}). Each system qubit is encoded in two
spatial modes representing $|0\rangle$ and $|1\rangle$, while the remaining
optical modes are single-rail~\cite{kang2026heralded}. In the uniform
setting, the $d$ outputs of a degree-$d$ ancilla panel carry equal
amplitudes $1/\sqrt{d}$. These output amplitudes are optimized to maximize
the success probability subject to normalization and preservation of the
target state. A sign-flipped edge weight is implemented by a $\pi$ phase
shifter on the corresponding output.

All outputs of each photon-subtraction-node panel are monitored by
detectors. A detection pattern specifies the detecting output in each
panel, with exactly one photon detected per panel.

\paragraph*{Symbolic validation and success probability.}
For each detection pattern, we calculate the output amplitudes of the
remaining qubits by evolving the optical creation operators and
projecting onto the corresponding detector outcome. A detection pattern
is called a heralding pattern if single-qubit phase shifts conditioned
on the detection outcome transform the corresponding normalized state
into the target state up to a global phase. The probability of each
pattern is the sum of the squared magnitudes of its output amplitudes.
The heralding success probability is the sum of these probabilities over
all heralding patterns.

\section{Circuits}
\label{appendix:circuits}

This appendix provides details of the heralded linear optical circuits
corresponding to the graph solutions in Sec.~\ref{sec:results}, including
their ancilla output amplitudes, generated states, and heralding success
probabilities. Each circuit is obtained using the graph-to-circuit
translation of Stage~3 (Appendix~\ref{appendix:stage3}). The output
amplitudes of each ancilla panel are listed in the top-to-bottom order
of the output rails in the corresponding figure.

\begin{figure*}[t]
    \centering
    \paperpanel{0.44\textwidth}{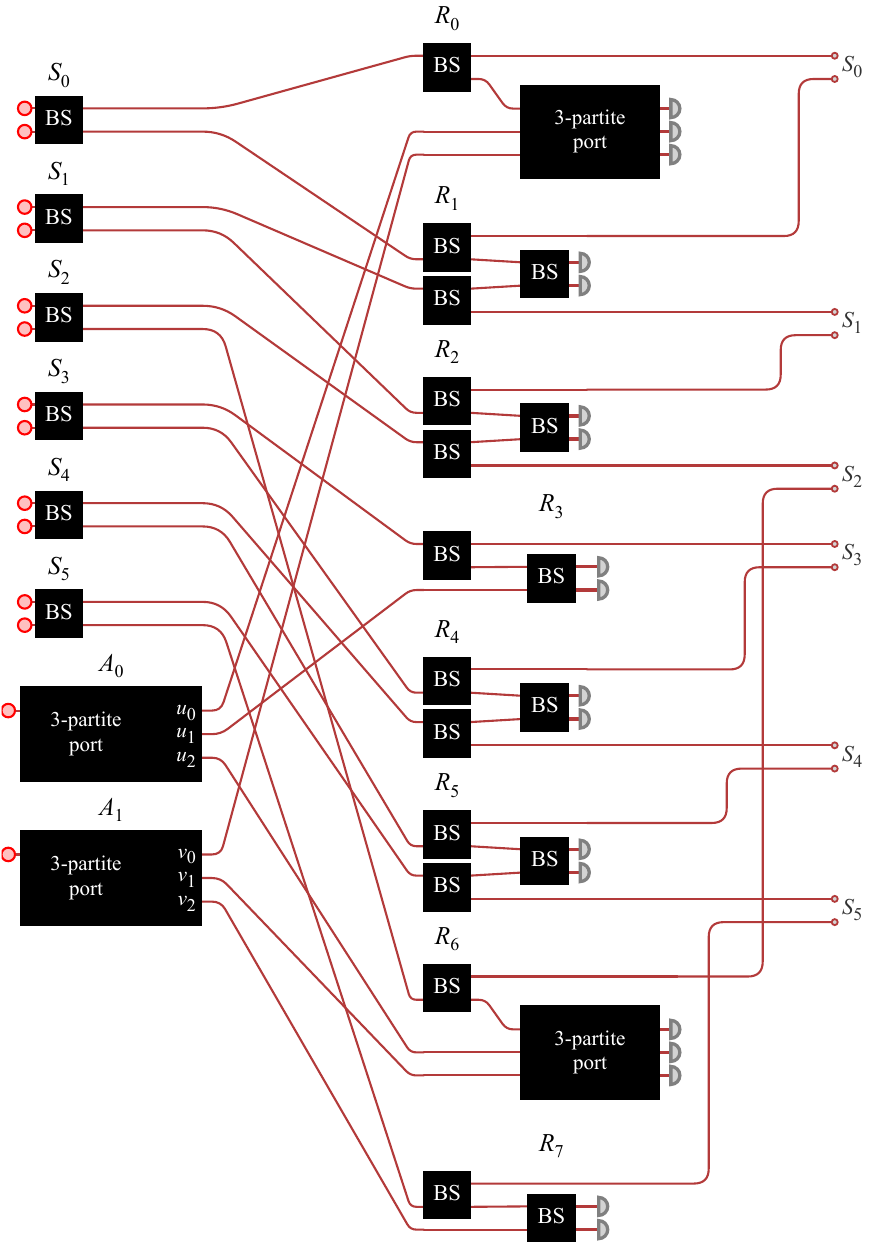}{a}\hfill
    \paperpanel{0.44\textwidth}{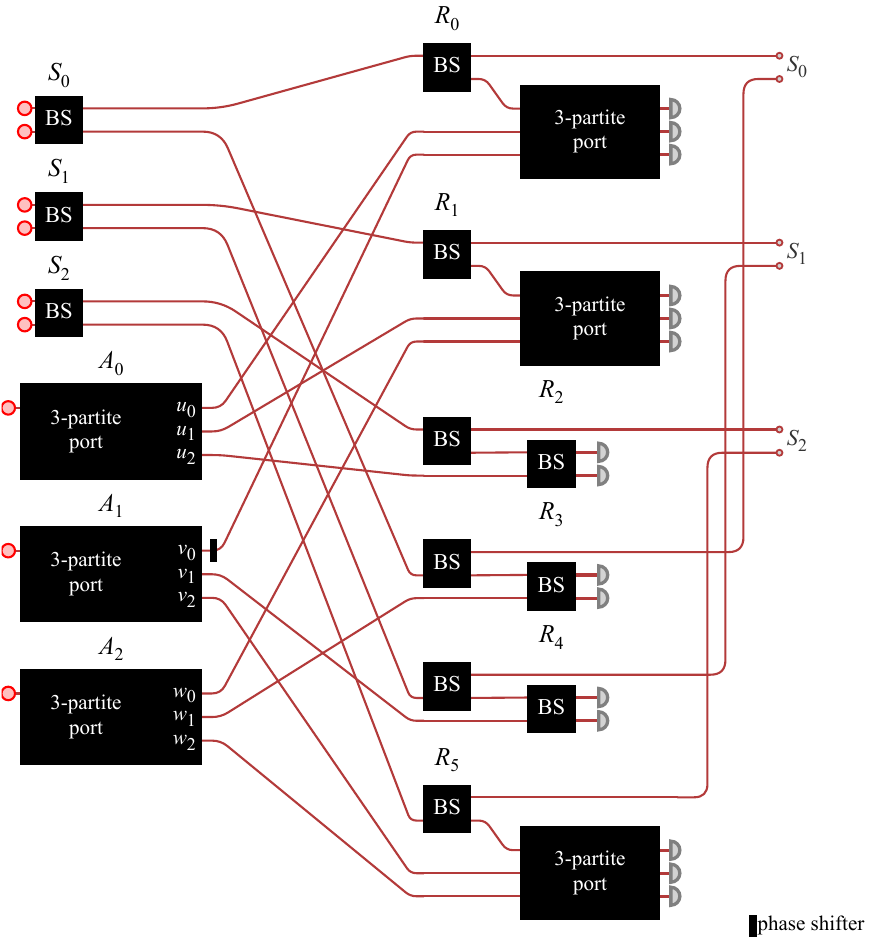}{b}
    \caption{Heralded circuits generating (a) $\a|0_L\> + \b|1_L\>$ for the
    $[[6,1,2]]$ loss-tolerant error-correcting
    code, translated from the graph of Fig.~\ref{fig:qec}(b), and (b) the
    $N=3$ amplitude-damping code, translated from the graph of
    Fig.~\ref{fig:qec}(d). The labels $u_k$, $v_k$, $w_k$ on
    the ancilla panel outputs indicate which amplitude of
    Eq.~\eqref{eq:qec_amplitude_vectors} or Eq.~\eqref{eq:adc3_amplitude_vectors}
    each rail carries; the $\pi$ phase shifter of circuit (b) acts on the
    $v_0$ rail.}
    \label{fig:qec6_adc3_circuits}
\end{figure*}

\begin{figure*}[t]
    \centering
    \paperpanel{0.44\textwidth}{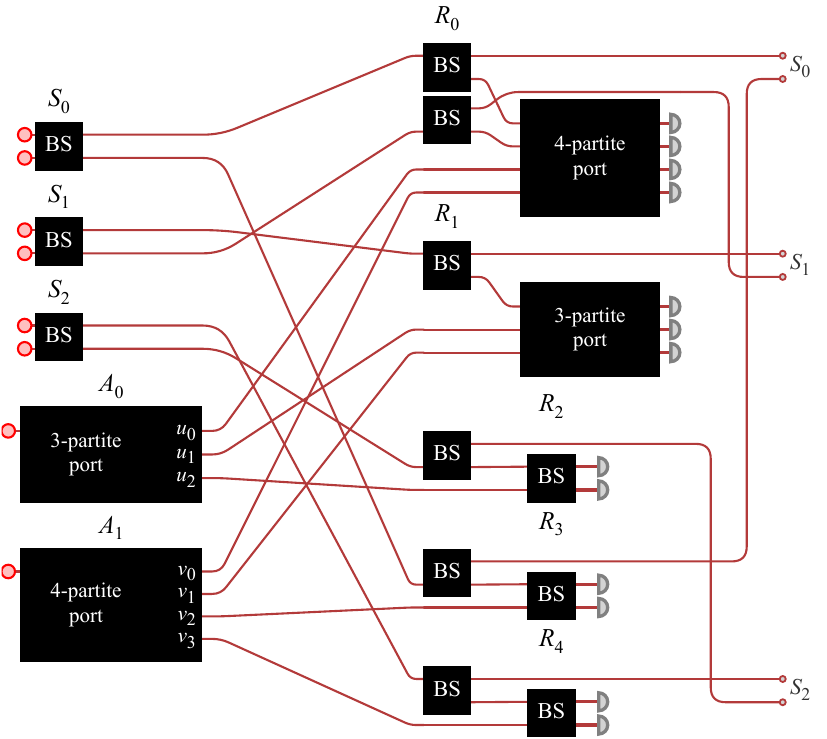}{a}\hfill
    \paperpanel{0.44\textwidth}{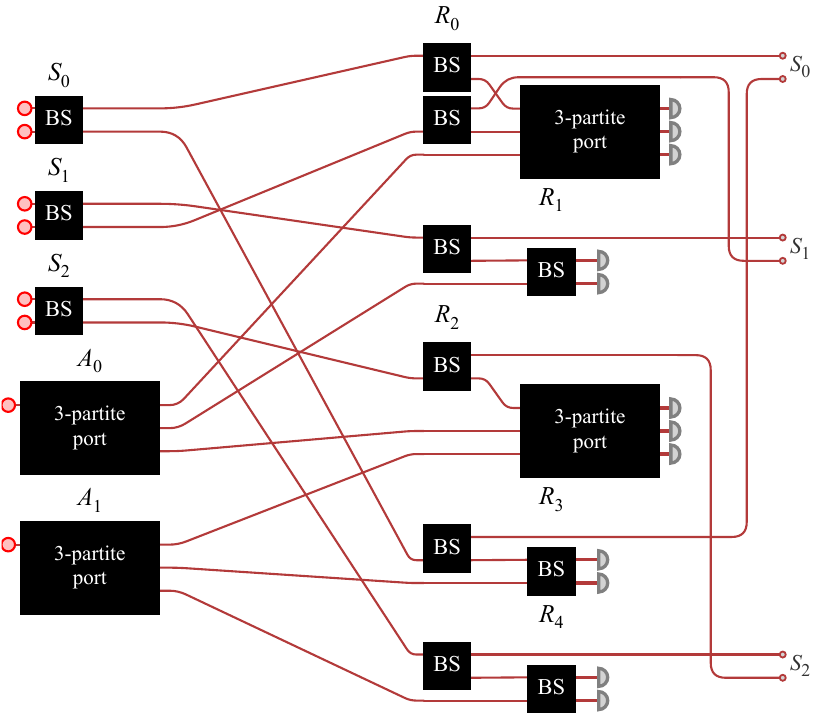}{b}
    \caption{Heralded circuits generating the $N=3$ Type 5 state,
translated from the graph solutions of Fig.~\ref{fig:type5}(a) and (b), respectively. Circuit (a) also realizes the arbitrary
tripartite family of Eq.~\eqref{eq:arbit_tripartite}, with the Type~5
state as its equal-coefficient case, whereas circuit (b) realizes only
the fixed Type~5 state.}
    \label{fig:type5_circuits}
\end{figure*}

\begin{figure}[t]
    \centering
    \includegraphics[width=\linewidth]{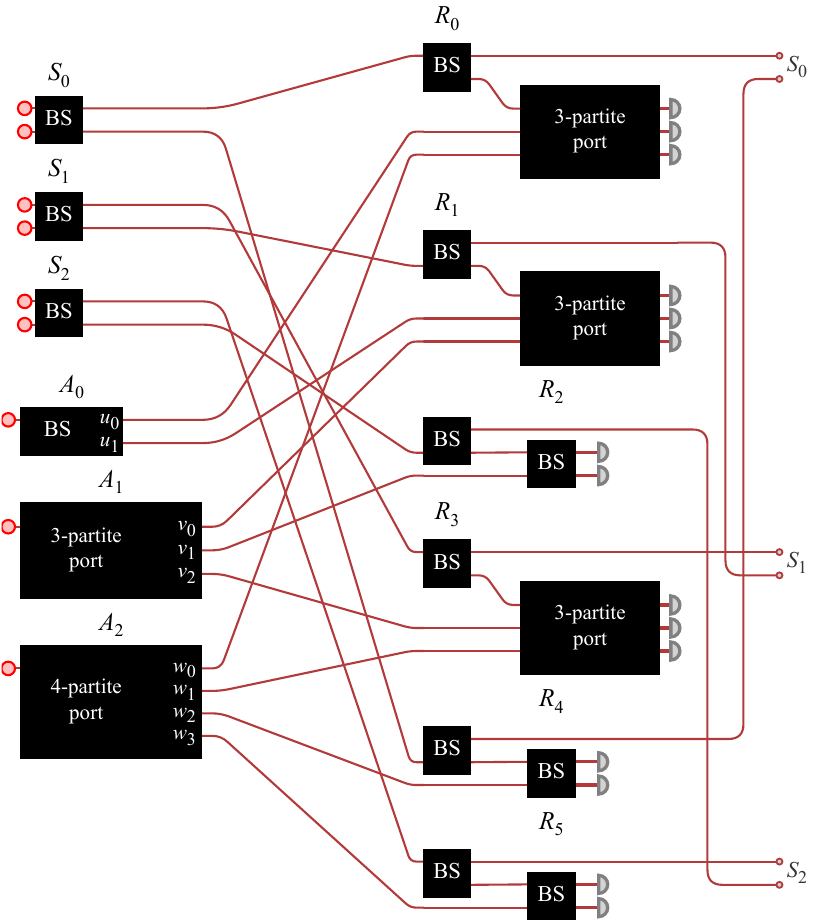}
    \caption{Heralded circuit generating the family of
    Eq.~\eqref{eq:arbit_tripartite}, translated from the $(3,3)$ graph of
    Fig.~\ref{fig:type5}(c). The rails
    $w_2$, $w_0$, $v_0$, $w_1$, and $v_2$ carry the coefficients $a$,
    $b\,e^{i\phi}$, $c$, $d$, and $e$, respectively.
    }
    \label{fig:type5_w2}
\end{figure}

\subsection{$N=3$ and $4$ hypergraph magic state circuits}
\label{appendix:hypergraph_4}

 The $N=3$ magic circuit of
Fig.~\ref{fig:magic3_circuit} has success probability
$P_{\text{succ}}=1/128$  with the
output amplitudes $A_0=\tfrac{1}{2}(1,\,\sqrt{2},\,1)$ and
$A_1=\tfrac{1}{2}(1,\,1,\,\sqrt{2})$.

For $N=4$, the graph solutions of Fig.~\ref{fig:magic_N}(b) and (c) translate into the
circuits of Fig.~\ref{fig:magic4_circuits}. The output amplitudes of
the ancilla panels are
\begin{align}
    &A_0=(0.569,\,0.593,\,0.569),\nn \\
    &A_1=(0.478,\,0.737,\,0.478),\nn \\
    &A_2=(0.420,\,0.420,\,0.438,\,0.676)
\end{align}
for circuit (a) and
\begin{align}
    &A_0=\tfrac{1}{2}\,(1,\,\sqrt{2},\,1),\qquad
    A_1=\tfrac{1}{2\sqrt{2}}\,(1,\,2,\,1,\,\sqrt{2}),\nn \\
    &A_2=\tfrac{1}{2\sqrt{2}}\,(1,\,1,\,\sqrt{2},\,2)
\end{align}
for circuit (b). Both circuits generate the $N=4$ hypergraph magic
state with $P_{\text{succ}} \approx 3.51\times 10^{-4}$ for circuit (a) and
$P_{\text{succ}} = 1/49152$ for circuit (b).

\subsection{$[[6,1,2]]$ code circuit}
\label{appendix:qec6}

The $[[6,1,2]]$ graph solution of Fig.~\ref{fig:qec}(b) translates into
the circuit of Fig.~\ref{fig:qec6_adc3_circuits}(a).
For a normalized target
($|\a|^2+|\b|^2=1$), the output amplitudes of the two ancilla panels
are
\begin{align}\label{eq:qec_amplitude_vectors}
    A_0&=(u_0,u_1,u_2)=\frac{1}{\sqrt{2}}\,(\b,\,1,\,\a), \nn \\
    A_1&=(v_0,v_1,v_2)=\frac{1}{\sqrt{2}}\,(\a,\,\b,\,1).
\end{align}
A relative sign between $\a$ and $\b$ is realized as a
$\pi$ phase shifter on the rails carrying $\b$ ($u_0$ and $v_1$). 

\subsection{$N=3$ amplitude-damping code circuit}\label{appendix:N3_ADC}

The circuit corresponding to the graph of Fig.~\ref{fig:qec}(d) is
shown in Fig.~\ref{fig:qec6_adc3_circuits}(b).
The optimized ancilla-panel output amplitudes for the maximal success probability~\eqref{eq:adc3_closed_form}, specified before the
indicated $\pi$ phase shifter on $v_0$, are
\begin{align}\label{eq:adc3_amplitude_vectors}
    A_0&=(u_0,u_1,u_2)=(g_\star,\,g_\star,\,1)/\sqrt{2g_\star^2+1},\nn \\
    A_1&=(v_0,v_1,v_2)=A_2=(w_0,w_1,w_2) \nn \\
    &=(xg_\star,\,g_\star,\,1)/\sqrt{(\chi+1)g_\star^2+1},
\end{align}
where $\chi \equiv \sqrt{3}|\b/\a|$ and
$g_\star^2\equiv(1+\sqrt{4\chi+5})/[2(\chi+1)]$.
For $\a\ne0$, choose a complex parameter $x$ satisfying
$x^2=-\sqrt{3}\,\b/\a$, so that $|x|^2=\chi$.

\subsection{General three-qubit state circuits}
\label{appendix:type5}

The circuit in Fig.~\ref{fig:type5_circuits}(a) generates the general three-qubit family of Eq.~\eqref{eq:arbit_tripartite}. Define $r=\sqrt{a^2+b^2+d^2}$ and $s=\sqrt{c^2+e^2}$. The normalized ancilla output amplitudes are
\begin{align}\label{eq:type5_general_vectors}
A_0&=(u_0,u_1,u_2)
=\frac{(c,e,t)}{\sqrt{s^2+t^2}},\nn \\
A_1&=(v_0,v_1,v_2,v_3)
=\frac{(b e^{i\phi},d,a,t)}{\sqrt{r^2+t^2}}.
\end{align}
Here $t>0$ is free. The phase $e^{i\phi}$ on $v_0$ affects only $|100\>$. The five perfect-matching products, ordered as in Eq.~\eqref{eq:arbit_tripartite}, are
\begin{align}\label{eq:type5_pm_products}
&(u_2v_2,u_2v_0,u_0v_3,u_2v_1,u_1v_3)\nn \\
&\qquad=
\frac{t}{\sqrt{(r^2+t^2)(s^2+t^2)}}
(a,b e^{i\phi},c,d,e).
\end{align}
Varying $t$ changes only the common prefactor.

Let $h_{R_k}$ denote the clicked detector in panel $R_k$, numbered from the top. A detection pattern has $h_{R_0}\in\{0,1,2,3\}$, $h_{R_1}\in\{0,1,2\}$, and $h_{R_2},h_{R_3},h_{R_4}\in\{0,1\}$. All 96 patterns are equally probable. For $a,\ldots,e>0$, click-conditioned single-qubit phase corrections give
\begin{align}\label{eq:type5_phase_h}
&a|000\>+b e^{i\phi_h}|100\>
+c|101\>+d|110\>+e|111\>,\nn \\
&\phi_h=\phi+\frac{\pi}{2}h_{R_0}
-\frac{2\pi}{3}h_{R_1}\pmod{2\pi}.
\end{align}
These patterns form 12 equally probable phase classes, each containing eight detection patterns (Table~\ref{tab:type5_phase_classes}); only $h_{R_0}=h_{R_1}=0$ gives the prescribed $\phi$.

\begin{table}[t]
\caption{Residual phase $\phi_h-\phi$ for Fig.~\ref{fig:type5_circuits}(a).
Each entry represents eight detection patterns; the eight patterns
in the zero entry are heralding patterns.}
\label{tab:type5_phase_classes}
\begin{ruledtabular}
\begin{tabular}{ccccc}
 & $h_{R_0}=0$ & $h_{R_0}=1$ & $h_{R_0}=2$ & $h_{R_0}=3$ \\
\hline
$h_{R_1}=0$ & $0$      & $\pi/2$    & $\pi$    & $3\pi/2$ \\
$h_{R_1}=1$ & $4\pi/3$ & $11\pi/6$  & $\pi/3$  & $5\pi/6$ \\
$h_{R_1}=2$ & $2\pi/3$ & $7\pi/6$   & $5\pi/3$ & $\pi/6$  \\
\end{tabular}
\end{ruledtabular}
\end{table}

Summing $p_h=t^2/[6144(r^2+t^2)(s^2+t^2)]$ over  all the  heralding patterns gives the total success probability~\eqref{eq:type5_closed_form}. 
If any of $(b,c,d,e)$ vanishes, the residual phase is removable and the success probability is equal to~\eqref{eq:type5_closed_form}.
If only $a$ vanishes, only eight patterns are suitable heralding patterns. For $r=0$ or $s=0$, we can take the $t\to0^+$ limit of Eq.~\eqref{eq:type5_general_vectors}. All 96 patterns are then heralding patterns, giving $P_{\text{succ}}=1/64$.
For example,  by setting $a=c=d=e=1/2$ and $b=0$, followed by Pauli-$X$ on all three qubits, the circuit generates the $N=3$ magic state with $P_{\text{succ}}=1/128$, the same as for the circuit in Fig.~\ref{fig:magic3_circuit}.




The Type~5 circuit in Fig.~\ref{fig:type5_circuits} (b) uses $A_0=A_1=(1,1,1)/\sqrt3$ and accepts all 72 detection patterns, giving $P_{\text{succ}}=5/(2^6 3^2)$. 

The nine-photon circuit in Fig.~\ref{fig:type5_w2}, corresponding to the $(3,3)$ graph in Fig.~\ref{fig:type5}(c), also generates the general three-qubit family of Eq.~\eqref{eq:arbit_tripartite}. By introducing two free parameters $z,t>0$ and defining $s=\sqrt{c^2+e^2}$ and $r_z=\sqrt{z(a^2+b^2)+d^2}$, we express the normalized ancilla output amplitudes as
\begin{align}\label{eq:w2_general_vectors}
A_0&=(u_0,u_1)=\frac{(\sqrt z,1)}{\sqrt{1+z}},\nn \\
A_1&=(v_0,v_1,v_2)=\frac{(c,t,e)}{\sqrt{s^2+t^2}},\nn \\
A_2&=(w_0,w_1,w_2,w_3)=\frac{(\sqrt z\,b e^{i\phi},d,\sqrt z\,a,t)}{\sqrt{r_z^2+t^2}}.
\end{align}
\begin{samepage}
In the basis order of Eq.~\eqref{eq:arbit_tripartite}, the five perfect-matching products satisfy
\begin{align}\label{eq:w2_pm_products}
&(u_1v_1w_2,u_1v_1w_0,u_0v_0w_3,u_0v_1w_1,u_0v_2w_3)\nn \\
&\qquad\propto(a,b e^{i\phi},c,d,e).
\end{align}
\end{samepage}

This circuit has 216 equally probable detection patterns, with $h_{R_0},h_{R_1},h_{R_3}\in\{0,1,2\}$ and $h_{R_2},h_{R_4},h_{R_5}\in\{0,1\}$. Click-conditioned single-qubit phase corrections give Eq.~\eqref{eq:arbit_tripartite} with $\phi$ replaced by
\begin{align}\label{eq:w2_phase_h}
\phi_h=\phi-\frac{2\pi}{3}\bigl(2h_{R_0}+h_{R_1}+h_{R_3}\bigr)\pmod{2\pi}.
\end{align}
The three equally probable phase classes each contain 72 patterns. For $(2h_{R_0}+h_{R_1}+h_{R_3})\bmod3=0,1,2$, the residual phases are $0,4\pi/3,2\pi/3$, respectively. 

\begin{figure}[t]
    \centering
    \includegraphics[width=1\linewidth]{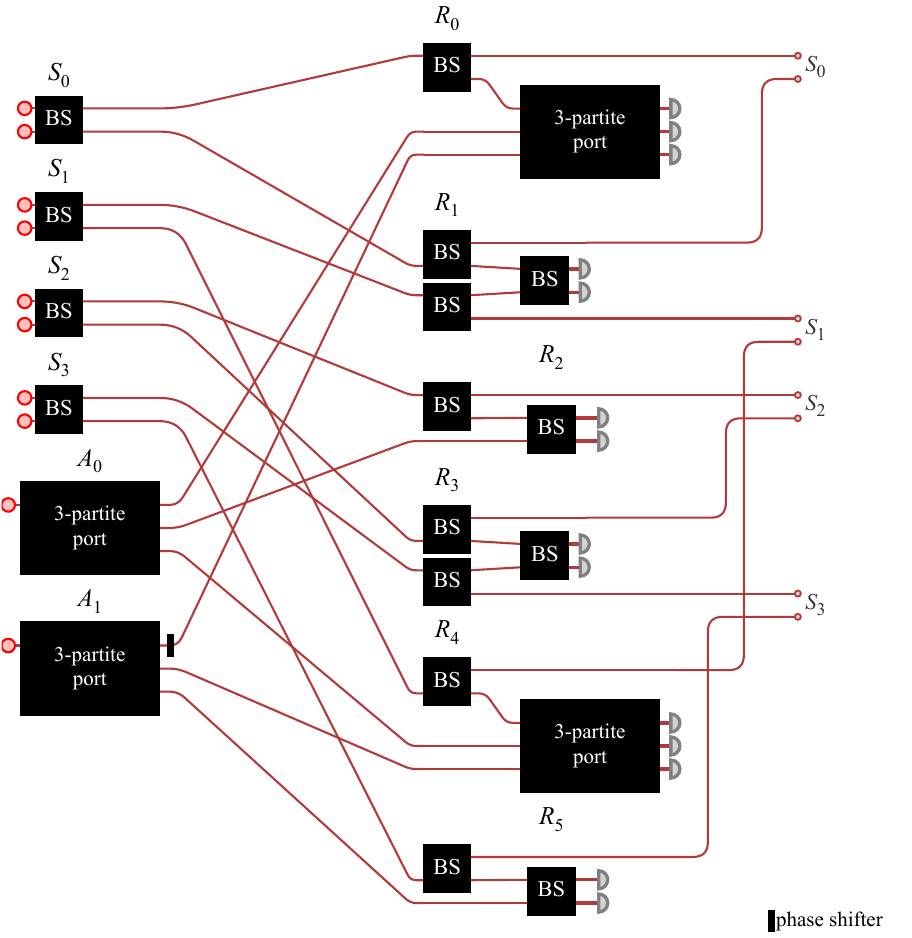}
    \caption{Heralded circuit generating the $N=4$ cluster state,
    translated from the graph of Fig.~\ref{fig:cluster4}(a). The vertical mark
    on the first output rail of $A_1$ denotes a $\pi$ phase shifter.}
    \label{fig:cluster4_circuit}
\end{figure}

For a given $z$, optimizing $t$ gives $t^2=s r_z$. Then the total success probability summed over all the heralding patterns
are given by Eq.~\eqref{eq:w2_general_success}, as a function of $z$. We can control $z$ in the circuit for the best success probability. 

For the Type~5 state, we can directly check that $z=(1+\sqrt3)/2$ gives the best success probability 
$P_{\text{succ}}=\frac{5(2\sqrt3-3)}{1728}\approx1.34\times10^{-3}$,
about twice of that Fig.~\ref{fig:type5_circuits}(a), using one additional photon.

\subsection{$N=4$ cluster state circuit}
\label{appendix:cluster4}

The graph solution of Fig.~\ref{fig:cluster4}(a) translates into the
circuit of Fig.~\ref{fig:cluster4_circuit}. The output amplitudes of the ancilla panels are
\begin{align}
    A_0=\tfrac{1}{2}\,(1,\,\sqrt{2},\,1),\qquad
    A_1=\tfrac{1}{2}\,(1,\,1,\,\sqrt{2})
\end{align}
and the $\pi$ phase shifter on the marked rail of $A_1$ rotates the phase
of the $|1111\>$ component.

A previously reported scheme generates the $N=4$ cluster state using 9 photons with a success probability of $2.05\times10^{-3}$, but requires 45 beam splitters~\cite{hartnett2026automated}. 
Our scheme requires only 16 beam splitters, two asymmetric three-partite multiports, and two symmetric three-partite multiports. Since an asymmetric and a symmetric three-partite multiport can be decomposed into two and three beam splitters, respectively~\cite{reck1994experimental,clements2016optimal}, our scheme consists of 26 beam splitters after this decomposition.

\bibliographystyle{unsrt}
\bibliography{Auto_EPM}

\begin{thebibliography}{10}

\bibitem{horodecki2009quantum}
Ryszard Horodecki, Pawe{\l} Horodecki, Micha{\l} Horodecki, and Karol Horodecki.
\newblock Quantum entanglement.
\newblock {\em Reviews of Modern Physics}, 81(2):865--942, 2009.

\bibitem{pan2012multiphoton}
Jian-Wei Pan, Zeng-Bing Chen, Chao-Yang Lu, Harald Weinfurter, Anton Zeilinger, and Marek {\.Z}ukowski.
\newblock Multiphoton entanglement and interferometry.
\newblock {\em Reviews of Modern Physics}, 84(2):777--838, 2012.

\bibitem{walter2016multipartite}
Michael Walter, David Gross, and Jens Eisert.
\newblock Multipartite entanglement.
\newblock {\em Quantum Information: From Foundations to Quantum Technology Applications}, pages 293--330, 2016.

\bibitem{kok2007linear}
Pieter Kok, William~J Munro, Kae Nemoto, Timothy~C Ralph, Jonathan~P Dowling, and Gerard~J Milburn.
\newblock Linear optical quantum computing with photonic qubits.
\newblock {\em Reviews of Modern Physics}, 79(1):135--174, 2007.

\bibitem{papp2009characterization}
Scott~B Papp, Kyung~Soo Choi, Hui Deng, Pavel Lougovski, SJ~Van~Enk, and HJ~Kimble.
\newblock Characterization of multipartite entanglement for one photon shared among four optical modes.
\newblock {\em Science}, 324(5928):764--768, 2009.

\bibitem{barz2010heralded}
Stefanie Barz, Gunther Cronenberg, Anton Zeilinger, and Philip Walther.
\newblock Heralded generation of entangled photon pairs.
\newblock {\em Nature Photonics}, 4(8):553--556, 2010.

\bibitem{li2021heralded}
Jin-Peng Li, Xuemei Gu, Jian Qin, Dian Wu, Xiang You, Hui Wang, Christian Schneider, Sven H{\"o}fling, Yong-Heng Huo, Chao-Yang Lu, et~al.
\newblock Heralded nondestructive quantum entangling gate with single-photon sources.
\newblock {\em Physical Review Letters}, 126(14):140501, 2021.

\bibitem{chin2024shortcut}
Seungbeom Chin, Yong-Su Kim, and Marcin Karczewski.
\newblock Shortcut to multipartite entanglement generation: A graph approach to boson subtractions.
\newblock {\em npj Quantum Information}, 10(1):67, 2024.

\bibitem{chin2024heralded}
Seungbeom Chin, Marcin Karczewski, and Yong-Su Kim.
\newblock Heralded optical entanglement generation via the graph picture of linear quantum networks.
\newblock {\em Quantum}, 8:1572, 2024.

\bibitem{forbes2025heralded}
Imogen Forbes, Farzad Ghafari, Edward~CR Deacon, Sukhjit~P Singh, Emilien Lavie, Patrick Yard, Reece~D Shaw, Anthony Laing, and Nora Tischler.
\newblock Heralded generation of entanglement with photons.
\newblock {\em Reports on Progress in Physics}, 88(8):086002, 2025.

\bibitem{bouwmeester1999observation}
Dik Bouwmeester, Jian-Wei Pan, Matthew Daniell, Harald Weinfurter, and Anton Zeilinger.
\newblock Observation of three-photon {G}reenberger-{H}orne-{Z}eilinger entanglement.
\newblock {\em Physical Review Letters}, 82(7):1345, 1999.

\bibitem{pan2001experimental}
Jian-Wei Pan, Matthew Daniell, Sara Gasparoni, Gregor Weihs, and Anton Zeilinger.
\newblock Experimental demonstration of four-photon entanglement and high-fidelity teleportation.
\newblock {\em Physical Review Letters}, 86(20):4435, 2001.

\bibitem{krenn2021conceptual}
Mario Krenn, Jakob~S Kottmann, Nora Tischler, and Al{\'a}n Aspuru-Guzik.
\newblock Conceptual understanding through efficient automated design of quantum optical experiments.
\newblock {\em Physical Review X}, 11(3):031044, 2021.

\bibitem{cervera2022design}
Alba Cervera-Lierta, Mario Krenn, and Al{\'a}n Aspuru-Guzik.
\newblock Design of quantum optical experiments with logic artificial intelligence.
\newblock {\em Quantum}, 6:836, 2022.

\bibitem{ruiz2023digital}
Carlos Ruiz-Gonzalez, S{\"o}ren Arlt, Jan Petermann, Sharareh Sayyad, Tareq Jaouni, Ebrahim Karimi, Nora Tischler, Xuemei Gu, and Mario Krenn.
\newblock Digital discovery of 100 diverse quantum experiments with {P}y{T}heus.
\newblock {\em Quantum}, 7:1204, 2023.

\bibitem{hartnett2026automated}
Gavin~S Hartnett, Dave Kielpinski, Smarak Maity, Pranav~S Mundada, Yuval Baum, and Michael~R Hush.
\newblock Automated discovery of heralded ballistic graph state generators for fusion-based photonic quantum computation.
\newblock {\em Physical Review A}, 113(4):042608, 2026.

\bibitem{chin2021graph}
Seungbeom Chin, Yong-Su Kim, and Sangmin Lee.
\newblock Graph picture of linear quantum networks and entanglement.
\newblock {\em Quantum}, 5:611, 2021.

\bibitem{chin2024exponentially}
Seungbeom Chin, Junghee Ryu, and Yong-Su Kim.
\newblock Exponentially enhanced scheme for the heralded qudit {G}reenberger-{H}orne-{Z}eilinger state in linear optics.
\newblock {\em Physical Review Letters}, 133(25):253601, 2024.

\bibitem{chin2026efficient}
Seungbeom Chin and William~J. Munro.
\newblock Efficient graph state generation in linear optics.
\newblock {\em Quantum}, 10:2189, 2026.

\bibitem{kang2026heralded}
Minhyeok Kang, Jaehee Kim, William~J Munro, Seungbeom Chin, and Joonsuk Huh.
\newblock Heralded linear optical generation of {D}icke states.
\newblock {\em New Journal of Physics}, 28(5):054501, 2026.

\bibitem{nielsen2010quantum}
Michael~A Nielsen and Isaac~L Chuang.
\newblock {\em Quantum computation and quantum information}.
\newblock Cambridge University Press, 2010.

\bibitem{kraus2010localprl}
Barbara Kraus.
\newblock Local unitary equivalence of multipartite pure states.
\newblock {\em Physical Review Letters}, 104(2):020504, 2010.

\bibitem{kraus2010localpra}
Barbara Kraus.
\newblock Local unitary equivalence and entanglement of multipartite pure states.
\newblock {\em Physical Review A}, 82(3):032121, 2010.

\bibitem{browne2005resource}
Daniel~E Browne and Terry Rudolph.
\newblock Resource-efficient linear optical quantum computation.
\newblock {\em Physical Review Letters}, 95(1):010501, 2005.

\bibitem{varnava2006loss}
Michael Varnava, Daniel~E Browne, and Terry Rudolph.
\newblock Loss tolerance in one-way quantum computation via counterfactual error correction.
\newblock {\em Physical Review Letters}, 97(12):120501, 2006.

\bibitem{bravyi2005universal}
Sergey Bravyi and Alexei Kitaev.
\newblock Universal quantum computation with ideal {C}lifford gates and noisy ancillas.
\newblock {\em Physical Review A—Atomic, Molecular, and Optical Physics}, 71(2):022316, 2005.

\bibitem{veitch2014resource}
Victor Veitch, SA~Hamed~Mousavian, Daniel Gottesman, and Joseph Emerson.
\newblock The resource theory of stabilizer quantum computation.
\newblock {\em New Journal of Physics}, 16(1):013009, 2014.

\bibitem{rossi2013quantum}
Matteo Rossi, Marcus Huber, Dagmar Bru{\ss}, and Chiara Macchiavello.
\newblock Quantum hypergraph states.
\newblock {\em New Journal of Physics}, 15(11):113022, 2013.

\bibitem{huang2024demonstration}
Jieshan Huang, Xudong Li, Xiaojiong Chen, Chonghao Zhai, Yun Zheng, Yulin Chi, Yan Li, Qiongyi He, Qihuang Gong, and Jianwei Wang.
\newblock Demonstration of hypergraph-state quantum information processing.
\newblock {\em Nature Communications}, 15(1):2601, 2024.

\bibitem{poderini2026quantum}
Davide Poderini, Dagmar Bru{\ss}, and Chiara Macchiavello.
\newblock Quantum hypergraph states: A review.
\newblock {\em Reports on Progress in Physics}, 89(6):066001, 2026.

\bibitem{ralph2007efficient}
TC~Ralph, KJ~Resch, and Alexei Gilchrist.
\newblock Efficient {T}offoli gates using qudits.
\newblock {\em Physical Review A—Atomic, Molecular, and Optical Physics}, 75(2):022313, 2007.

\bibitem{uskov2009maximal}
Dmitry~B Uskov, Lev Kaplan, A~Matthew Smith, Sean~D Huver, and Jonathan~P Dowling.
\newblock Maximal success probabilities of linear-optical quantum gates.
\newblock {\em Physical Review A—Atomic, Molecular, and Optical Physics}, 79(4):042326, 2009.

\bibitem{leung1997approximate}
Debbie~W Leung, Michael~A Nielsen, Isaac~L Chuang, and Yoshihisa Yamamoto.
\newblock Approximate quantum error correction can lead to better codes.
\newblock {\em Physical Review A}, 56(4):2567, 1997.

\bibitem{dutta2026smallest}
Sourav Dutta, Aditya Jain, and Prabha Mandayam.
\newblock Smallest quantum codes for amplitude-damping noise.
\newblock {\em Physical Review Research}, 8(3):L032004, 2026.

\bibitem{acin2000generalized}
Antonio Ac{\'\i}n, A~Andrianov, L~Costa, E~Jan{\'e}, JI~Latorre, and Rolf Tarrach.
\newblock Generalized {S}chmidt decomposition and classification of three-quantum-bit states.
\newblock {\em Physical Review Letters}, 85(7):1560, 2000.

\bibitem{blasiak2022arbitrary}
Pawel Blasiak, Ewa Borsuk, and Marcin Markiewicz.
\newblock Arbitrary entanglement of three qubits via linear optics.
\newblock {\em Scientific Reports}, 12(1):21596, 2022.

\bibitem{raussendorf2001one}
Robert Raussendorf and Hans~J Briegel.
\newblock A one-way quantum computer.
\newblock {\em Physical Review Letters}, 86(22):5188, 2001.

\bibitem{azuma2015all}
Koji Azuma, Kiyoshi Tamaki, and Hoi-Kwong Lo.
\newblock All-photonic quantum repeaters.
\newblock {\em Nature {C}ommunications}, 6(1):6787, 2015.

\bibitem{hartmann2007weighted}
Lorenz Hartmann, J~Calsamiglia, W~D{\"u}r, and HJ~Briegel.
\newblock Weighted graph states and applications to spin chains, lattices and gases.
\newblock {\em Journal of Physics B: Atomic, Molecular and Optical Physics}, 40(9):S1--S44, 2007.

\bibitem{plato2008random}
A~Douglas~K Plato, Oscar~C Dahlsten, and Martin~B Plenio.
\newblock Random circuits by measurements on weighted graph states.
\newblock {\em Physical Review A—Atomic, Molecular, and Optical Physics}, 78(4):042332, 2008.

\bibitem{reck1994experimental}
Michael Reck, Anton Zeilinger, Herbert~J Bernstein, and Philip Bertani.
\newblock Experimental realization of any discrete unitary operator.
\newblock {\em Physical Review Letters}, 73(1):58--61, 1994.

\bibitem{clements2016optimal}
William~R Clements, Peter~C Humphreys, Benjamin~J Metcalf, W~Steven Kolthammer, and Ian~A Walmsley.
\newblock Optimal design for universal multiport interferometers.
\newblock {\em Optica}, 3(12):1460--1465, 2016.

\end{thebibliography}

\end{document}